\documentclass[superscriptaddress, a4paper,twocolumn,11pt,published,accepted=2020-04-04]{quantumarticle}
\pdfoutput=1

\usepackage[english]{babel}
\usepackage[utf8]{inputenc}

\usepackage[sort&compress,numbers]{natbib}
\usepackage{amsthm, 
color, 
mathtools, 
amsmath, 
amssymb,
setspace, 
bbm, 
tensor, 
subfigure, 
float,
mathtools, 
placeins, 
mathrsfs}
\usepackage[dvipsnames]{xcolor}
\usepackage{graphicx}
\usepackage[unicode=true, breaklinks=false, pdfborder={0 0 1}, backref=false, colorlinks=true, linkcolor=blue, citecolor=blue]{hyperref}
\usepackage{cancel,bm}

\let\oldsqrt\sqrt
\def\sqrt{\mathpalette\DHLhksqrt}
\def\DHLhksqrt#1#2{%
\setbox0=\hbox{$#1\oldsqrt{#2\,}$}\dimen0=\ht0
\advance\dimen0-0.2\ht0
\setbox2=\hbox{\vrule height\ht0 depth -\dimen0}%
{\box0\lower0.4pt\box2}}

\newcommand{\tr}{\operatorname{tr}}

\newcommand{\Bcal}{\mathcal{B}}
\newcommand{\Ecal}{\mathcal{E}}
\newcommand{\Fcal}{\mathcal{F}}
\newcommand{\Hcal}{\mathcal{H}}
\newcommand{\Ical}{\mathcal{I}}
\newcommand{\Mcal}{\mathcal{M}}
\newcommand{\Tcal}{\mathcal{T}}
\newcommand{\Ucal}{\mathcal{U}}

\newcommand{\Lcal}{\mathcal{L}}

\newcommand{\Jcal}{\mathcal{J}}
\newcommand{\Pcal}{\mathcal{P}}
\newcommand{\Pprob}{\mathbb{P}}

\newcommand{\Kcal}{\mathcal{K}}

\newcommand{\inp}{\texttt{i}}
\newcommand{\out}{\texttt{o}}

\DeclareMathAlphabet\mathbfcal{OMS}{cmsy}{b}{n}

\def\bra#1{\mathinner{\langle{#1}|}}

\def\ket#1{\mathinner{|{#1}\rangle}}

\def\braket#1{\mathinner{\langle{#1}\rangle}}

\newcommand*\xbar[1]{%
   \hbox{%
     \vbox{%
       \hrule height 0.5pt 
       \kern0.5ex
       \hbox{%
         \kern-0.2em
         \ensuremath{#1}%
         \kern-0.0em
       }%
     }%
   }%
} 

\def\BraVert{\egroup\,\mid\,\bgroup}

\def\ketbra#1#2{\ket{#1\vphantom{#2}}\!\bra{#2\vphantom{#1}}}

\def\bra#1{\mathinner{\langle{#1}|}}

\def\ket#1{\mathinner{|{#1}\rangle}}

\def\braket#1{\mathinner{\langle{#1}\rangle}}

\usepackage{bbm, braket}
\usepackage[mathscr]{euscript}
\allowdisplaybreaks
\newtheorem*{theorem*}{Theorem}

\newtheorem{proposition}{Proposition}
\newtheorem{observation}{Observation}
\newtheorem*{remark*}{Remark}

\begin{document}

\title{Kolmogorov extension theorem for (quantum) causal modelling and general probabilistic theories}

\author{Simon Milz}
\email{simon.milz@oeaw.ac.at} 
\affiliation{Institute for Quantum Optics and Quantum Information, Austrian Academy of Sciences, Boltzmanngasse 3, 1090 Vienna, Austria}
\affiliation{School of Physics and Astronomy, Monash University, Clayton, Victoria 3800, Australia}
\orcid{0000-0002-6987-5513}

\author{Fattah Sakuldee}
\affiliation{International Centre for Theory of Quantum Technologies, University of Gda\'nsk, Wita Stwosza 63, 80-308 Gda\'nsk, Poland}
\affiliation{MU-NECTEC Collaborative Research Unit on Quantum Information,
Department of Physics, Faculty of Science, Mahidol University, Bangkok 10400, Thailand.}
\orcid{0000-0001-8756-7904}
    
\author{Felix A. Pollock}
\orcid{0000-0002-1483-5661}
\affiliation{School of Physics and Astronomy, Monash University, Clayton, Victoria 3800, Australia}

\author{Kavan Modi}
\orcid{0000-0002-2054-9901}
\affiliation{School of Physics and Astronomy, Monash University, Clayton, Victoria 3800, Australia}
\date{October 7, 2019}

\begin{abstract}
In classical physics, the Kolmogorov extension theorem lays the foundation for the theory of stochastic processes. It has been known for a long time that, in its original form, this theorem does not hold in quantum mechanics. More generally, it does not hold in any theory of stochastic processes -- classical, quantum or beyond -- that does not just describe passive observations, but allows for active interventions. Such processes form the basis of the study of causal modelling across the sciences, including in the quantum domain. To date, these frameworks
have lacked a conceptual underpinning similar to that provided by Kolmogorov’s theorem for
classical stochastic processes. We prove a generalized extension theorem that applies to \textit{all} theories of stochastic processes, putting them on equally firm mathematical ground as their classical counterpart. Additionally, we show that quantum causal modelling and quantum stochastic processes are equivalent. This provides the correct framework for the description of experiments involving continuous control, which play a crucial role in the development of quantum technologies. Furthermore, we show that the original extension theorem follows from the generalized one in the correct limit, and elucidate how a comprehensive understanding of general stochastic processes allows one to unambiguously define the distinction between those that are classical and those that are quantum.
\end{abstract}

\maketitle
\section{Introduction}
\label{sec::Intro}
Stochastic processes are ubiquitous in nature. Their theory is used, among other applications, to model the stock market, predict the weather, describe transport processes in cells and understand the random motion of particles suspended in a fluid~\cite{schuss_theory_2009, liao_applied_2013}. Intuitively, when we speak of stochastic processes, we often mean joint probability distributions of random variables at a finite set of times: the probability for a stock to have prices $P_1, P_2$ and $P_3$ on three subsequent days, or the probability to find a particle undergoing Brownian motion in regions $R_1$ and $R_2$ when measuring its position at times $t_1$ and $t_2$. 

This finite description of stochastic processes is motivated by both experimental and mathematical considerations. On the experimental side, temporal resolution is generally limited and digital instruments always record a finite amount of data. Hence, the only accessible information we are left with is encoded in probability distributions with a finite number of arguments. On the mathematical side, it is much less cumbersome to model stochastic processes on discrete times -- for example, by defining transition probabilities $\mathbbm{P}(Y|X)$ between random variables at a fixed set of different times -- than modelling probability densities on the space of all possible `trajectories' of random variables. 

These motivations notwithstanding, the fundamental laws of physics are continuous in nature and one always implicitly assumes that there is an \textit{underlying} process that leads to the experimentally accessible finite distributions. Put differently, one assumes that there exists an infinite joint probability distribution that has \textit{all} the finite ones as marginals. For classical stochastic processes, these two points of view, the finite and the infinite one, are reconciled by the Kolmogorov extension theorem (\textbf{KET}), which lays bare the minimal requirements for the existence of an underlying process, given a family of measurement statistics for finite sets of times~\cite{kolmogorov_foundations_1956, feller_introduction_1968, breuer_theory_2007, tao_introduction_2011}. It bridges the gap between experimental reality and mathematically rigorous theoretical underpinnings and, as such, enables the definition of stochastic processes as the limit of finite -- and hence observable -- objects. Additionally, the KET enables the modelling of continuous processes based on finite probability distributions. As a consequence, in the classical setting, stochastic processes over a continuous set of times, and families of finite probability distributions are two sides of the same coin.

The validity of the KET hinges crucially on the fact that the statistics of observations at a time $t$ do not depend on the kind of measurements that were made at any time $t'<t$. Put differently, just like the Leggett-Garg inequalities for temporal correlations~\cite{leggett_quantum_1985,leggett_realism_2008,emary_leggettgarg_2014}, the KET is based on the assumption of noninvasive measurements and realism \textit{per se}. While the latter property means that any measurement merely reveals a well-defined pre-existing value, the former implies that said measurements can be carried out without disturbing the state of the measured system. For example, in a classical stochastic process, measuring the position of a particle undergoing Brownian motion, reveals pre-existing information, but does not actively change the state of the particle. Both of these conditions together ensure the existence of compatible measurement statistics for different sets of times, which form the basis for the derivation of the KET.

On the other hand, the assumptions of non-invasiveness or realism \textit{per se} are not fulfilled in many experimental scenarios, leading to a breakdown of the KET, at the cost of a clear connection between an underlying process and its finite time manifestations.
This is the case whenever an experimenter chooses to actively interfere with a process to uncover its causal structure or to investigate the reaction to different inputs. For example, instead of just observing the progress of a disease, a pharmacologist tries to find out how the course of a disease changes with the administration of certain drugs. More generally, agent based modelling investigates how systems behave when they can not only be monitored, but actively influenced~\cite{gilbert_agent-based_2007}. Experimental situations where interventions are actively used to uncover causal relations fall within the field of \textit{causal modelling}~\cite{pearl_causality_2009}. 

Interventions appear naturally in quantum mechanics, where measurements necessarily perturb a system's state; in fact, a complete description of quantum processes without interventions is not possible~\cite{milz_introduction_2017}. As in the classical case, interventions can also be used to actively probe the causal structure of a quantum process, and the description of quantum processes with interventions has been recently used to develop the field of \textit{quantum causal modelling}~\cite{1367-2630-18-6-063032,oreshkov_causal_2016, allen_quantum_2017}. Importantly, as in the case of classical processes with interventions, the invasiveness of measurements means that the KET does not hold for quantum processes~\cite{BreuerEA2016}. This is analogous to the violation of Leggett-Garg inequalities~\cite{leggett_quantum_1985,emary_leggettgarg_2014} in quantum mechanics.

The fundamental lack of an extension theorem in quantum mechanics (or any other theory with interventions) would be problematic for several reasons: Firstly, it would suggest a lack of consistency between descriptions of a process for different sets of times; for example, the description of a process for three times $t_1, t_2,$ and $t_3$ would \textit{not} already include the description of the process for the two times $t_1$ and $t_3$ only. In other words, we would need seven different independent descriptors for each of the seven subsets of times to describe all possible events! This lack of consistency would render the study of (quantum) causal models in multi-step experiments impossible; if local interventions lead to a completely different process, it is not meaningful to try to deduce causal relations by means of active manipulations of the system at hand. 

Furthermore, the present situation (without an extension theorem) implies an incompatibility between existing frameworks to describe processes with interventions (both classical and quantum) and the classical theory of stochastic processes, even though they should converge to the latter in the correct limit. This then suggests that the mere act of interacting with a system over time introduces a fundamental divide between the continuity of physical laws and the finite statistics that can be accessed in reality, thus begging the question: What do we generally mean by a (quantum) `stochastic process', and how can we reconcile causal modelling frameworks with the idea of an underlying process?

In this Paper, we answer these questions by generalizing the KET to the framework of (quantum) causal modelling, thus closing the apparent divide between the finite and the continuous point of view. To this end, in Sec.~\ref{sec::ClassCaus} we reiterate the relation between classical stochastic processes and classical causal modelling and show the breakdown of the KET when we allow for active interventions in Sec.~\ref{sec::KET_inter}. We analyze the quantum case in Sec.~\ref{sec::QuantStoch} and find that stochastic processes can only be defined properly by taking interventions into account. Consequently, the framework of quantum stochastic processes is equivalent to quantum causal modelling. In Sec.~\ref{subsec::GET}, we prove our main result, that the KET can be generalized to quantum stochastic processes, and this \textit{generalized extension theorem} (\textbf{GET}) reduces to the classical one in the correct limit. The breakdown of the KET is a breakdown of formalism only, not a fundamental property of quantum processes. Our generalized extension theorem provides an overarching theorem that puts all processes with interventions and, in particular, quantum processes on an equally sound footing as their classical counterpart.

We discuss the equivalence of quantum stochastic processes and quantum causal modelling in Sec.~\ref{subsec::QSP_QCM}. As a direct application, in Sec.~\ref{subsec::GET_KET}, we use the GET to provide a distinction between general, \textit{i.e.}, non-Markovian, classical  and  quantum  processes, as has been recently introduced in~\cite{smirne_coherence_2017} for the restricted case of processes without memory. While we phrase our results predominantly in the language of causal modelling, they apply to a wide range of current theories of quantum processes and beyond. The relation of our results to other frameworks, in particular to the work of Accardi, Frigerio and Lewis~\cite{accardi_quantum_1982}, is discussed in Sec.~\ref{sec::Relation}. We conclude the Paper in Sec.~\ref{sec::Concl}.

\section{Classical stochastic processes and Causal Modelling}
\label{sec::ClassCaus}

\subsection{Classical stochastic processes}
A classical stochastic process can be described by joint probability distributions $\Pprob_{\Lambda_k} (i_k,\dots,i_1):= \Pprob_{\Lambda_k}(\mathbf{i}_{\Lambda_k})$ of random variables that take values $\{i_\alpha\}$ at time $t_\alpha$~\cite{breuer_theory_2007}, where $\Lambda_k$ is a collection of times with cardinality $|\Lambda_k|=k$. For instance, for a $k$ step process,  the set of times could be $\Lambda_k =\left\{t_k,\dots,t_1\right\}$. We will employ the convention that subscripts signify the time as well as the particular value of the respective random variable. For example, $i_\alpha$ signifies a value of the random variable at time $t_\alpha$. 

The distribution $\Pprob_{\Lambda_k}(\mathbf{i}_ k)$ could express the probability for a particular length-$k$ sequence of heads and tails when flipping a coin, or the probability to find a particle undergoing Brownian motion at positions $\mathbf{i}_k$ when measuring it at times $\Lambda_k$. Importantly, this description of a stochastic process is sufficient to describe the behaviour on any subset of the times considered; for instance, the distribution over all but the $j$th time is found by marginalizing the larger distribution: $\Pprob_{\Lambda_k\setminus \{t_j\}}(i_k,\dots,i_{j+1},i_{j-1},\dots, i_1) = \sum_{i_j} \Pprob_{\Lambda_k}(\mathbf{i}_ k)$. This property implicitly assumes that there is only one instrument that is used to interrogate the system of interest, and this interrogation does not influence its state. Neither of these assumption are fulfilled in more general processes, such as the ones employed in causal modelling.  

\subsection{Classical causal modelling}
\begin{figure}[t]
\centering
\includegraphics[width=0.95\linewidth]{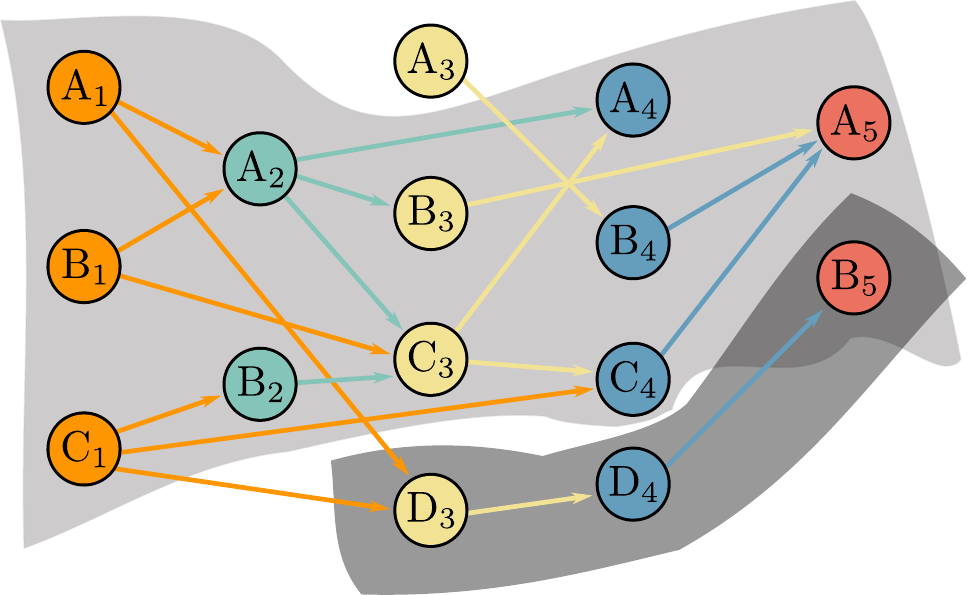}
\caption{\textit{(Quantum) Causal network.} Performing different interventions allows for the causal relations between events to be probed. For example, in the figure the event $B_1$ directly influences the events $C_3$ and $A_2$, while $A_3$ influences only $B_4$. Depending on the degrees of freedom that can be accessed by the experimenter, these causal relations can or cannot be detected. For example, the influence of $A_3$ on $B_4$ could not be discovered if only the degrees of freedom in the gray areas were experimentally accessible. Independent of the accessible degrees of freedom, the generalized extension theorem (GET) that we derive below holds for \textit{any} process. On the other hand, the statistics of events do in general not satisfy the requirements of the KET. For example, the events $D_3, D_4, B_5$ could be successive (\textit{e.g.}, at times $t_3,t_4$ and $t_5$) spin measurements in $z$-, $x$- and $z$-direction, respectively. Summing over the results of the spin measurement in $x$-direction at $t_4$ would not yield the correct probability distribution for two measurements in $z$-direction at $t_3$ and $t_5$ only (see also Sec.~\ref{subsec::KET_QM}). }
\label{fig::image3}
\end{figure}

Observing the statistics for measurement outcomes reveals correlations between events, but no information about causal relations. For instance, correlations of two events $A$ and $B$ could stem from $A$ influencing $B$, $B$ influencing $A$, or both $A$ and $B$ being influenced by an earlier event $C$~\cite{pearl_causality_2009,1367-2630-18-6-063032, allen_quantum_2017} (see Fig.~\ref{fig::image3}). Reiterating an example from Ref.~\cite{allen_quantum_2017}, events $A$ and $B$ could be the occurrence of sunburns and the sales of ice cream, respectively. While these two variables are highly correlated, this correlation alone would not fix a causal relation between them. Inferring the causal structure of a process is the aim of \textit{causal modelling}. Here, active interventions are used to uncover how different events can influence each other. In the example above, one could suspend the sale of ice cream to see how it affects the occurrence of sunburns, and would find out that ice cream sales have no direct effect on sunburns (and vice versa, as the correlations of ice cream sales and sunburns stem from a common cause, sunny weather, and not from any direct causal relation).  

Mathematically, causal modelling for $k$ events $A_k, \dots, A_1$ necessitates the collection of all joint probability distributions $\mathbbm{P}_{\Gamma_k}(i_k,\dots,i_1|j_k,\dots,j_1) := \mathbbm{P}_{\Gamma_k}(\mathbf{i}_{\Gamma_k}|\mathbfcal{J}_{\Gamma_k})$ to measure the outcomes $i_k,\dots,i_1$ given that the \textit{interventions} $j_k,\dots,j_1$ were performed. Here, $\Gamma_k$ is a set of labels for events; \textit{a priori}, there is no particular order imposed on the elements of $\Gamma_k$, and we use a different symbol for the set of event labels to distinguish it from the set of times $\Lambda_k$ used above. For example, $\Gamma_k$ could contain labels for different laboratories where experiments are performed. $\mathbfcal{J}_{\Gamma_k}$ are the \textit{instruments} that were used at each of the events; these can be seen as rules for how to intervene upon seeing a particular outcome (we will formalize the notion of an instrument in Sec.~\ref{sec::QuantStoch}).   For example, when investigating Brownian motion, an instrument could be a deterministic replacement rule: upon finding the particle at $i_a$ replace it by a particle at $i_a'$. It could also be probabilistic: upon finding the particle at $i_a$, with probability $p_{i_a'}$ replace it by a particle at $i_a'$. 

One possible instrument is the trivial idle instrument $\mathcal{J}_a = \mathrm{id}_a$,  the instrument that only measures the particle but doesn't change it. For classical stochastic processes, the corresponding joint distribution over outcomes can be thought of as the instrument-independent underlying distribution of the random variables describing the process:
\begin{gather}
    \label{eqn::idle}
    \mathbbm{P}_{\Gamma_k}(\mathbf{i}_{\Gamma_k}|\mathbf{id}_{\Gamma_k}) = \mathbbm{P}_{\Gamma_k}(\mathbf{i}_{\Gamma_k})\, ,
\end{gather}
where $\mathbf{id}_{\Gamma_k}$ denotes the idle instrument at each of the events in $\Gamma_k$. If we chose $\Gamma_k$ to be a set of times $\Lambda_k$, the right-hand side of Eq.~\eqref{eqn::idle} has the form of a $k$-step stochastic process. This directly leads to the following (well-known) observation:
\begin{observation}
\label{lem::ccm_csp}
Classical causal modelling contains classical stochastic processes as a special case. 
\end{observation}
As mentioned, this statement follows by choosing the set of events $\Gamma_k$ to be a set $\Lambda_k$ of ordered times and the instruments to be the idle instrument. We emphasize that causal modelling does not impose a temporal ordering \textit{per se}, but deduces the ordering of events from the obtained correlation functions (finding this order, or, more precisely, the underlying \textit{directed acyclic graph} (DAG) that defines the causal relations of the events, is the original aim of causal modelling~\cite{pearl_causality_2009,1367-2630-18-6-063032}). As neither the proof of the KET, nor the proof of the GET makes use of the notion of \textit{a priori} temporal ordering (see Sec.~\ref{subsec::GET} for a discussion), in what follows, we will drop the distinction between sets of labels $\Gamma_k$ and sets of times $\Lambda_k$. We now show that the introduction of interventions, that is crucial for the deduction of causal relations, leads to a breakdown of fundamental properties that are satisfied by classical stochastic processes.

\section{The Kolmogorov extension theorem and interventions}
\label{sec::KET_inter}
\subsection{The KET}
The KET is concerned with the question of which properties a family of finite joint probability distributions have to satisfy in order for an underlying process to exist. As such, it defines the classical notion of a stochastic process. In what follows, we will distinguish between stochastic processes on a finite number of times -- which are characterized by joint probability distributions with finitely many arguments -- and the \textit{underlying} stochastic process that leads to \textit{all} of these finite distributions.

As already mentioned, a classical stochastic process is described by a family of joint probability distributions $\Pprob_{\Lambda_k} (\mathbf{i}_{\Lambda_k})$ for different finite sets of times $\Lambda_k$. 
An underlying process on a set $\Lambda$ (finite, countably or uncountably infinite) is a joint probability distribution $\mathbbm{P}_{\Lambda}$, that has all finite ones as marginals. In detail, we have $\mathbbm{P}_{\Lambda_k}(\mathbf{i}_ {\Lambda_k}) = \sum_{\Lambda \setminus \Lambda_k}\mathbbm{P}_{\Lambda}(\mathbf{i}_\Lambda) := \Pprob_{\Lambda}^{|\Lambda_k}(\mathbf{i}_{\Lambda_k})$ for all $\Lambda_k \subseteq \Lambda$, where $\mathbf{i}_{\Lambda_k}$ is the subset of $\mathbf{i}_\Lambda$ corresponding to the times $\Lambda_k$, $\sum_{\Lambda\setminus \Lambda_k}$ denotes a sum over realizations of the random variables at all times that are part of $\Lambda \setminus \Lambda_k$ (\textit{i.e.}, all the times that lie in $\Lambda$ but not in $\Lambda_k$), and $\mathbbm{P}_{\Lambda}^{|\Lambda_k}$ denotes the restriction of $\mathbbm{P}_{\Lambda}$ to the times $\Lambda_k$. In the case where the set $\Lambda$ is infinite, the marginalization procedure can correspond to an integral over the times in $\Lambda\setminus \Lambda_k$ (though, to avoid introducing too much notation, we will still use $\sum_{\Lambda\setminus \Lambda_k}$ to represent it). For example, if the process we are interested in is the Brownian motion of a particle, $\mathbbm{P}_\Lambda$ would be the probability density of all possible trajectories that the particle could take in the time interval $\Lambda$, and all finite distributions could in principle be obtained from $\mathbbm{P}_\Lambda$.

If the finite joint probability distributions stem from an underlying process, it is easy to see that probability distributions for any two finite subsets of times $\Lambda_k \subseteq \Lambda_\ell \subseteq \Lambda$ fulfill a \emph{consistency condition} (or \textit{compatibility condition}) amongst each other, \textit{i.e.}, $\mathbbm{P}_{\Lambda_k}$ is a marginal of $\mathbbm{P}_{\Lambda_\ell}$. Expressed in the notation introduced above, we have $\mathbbm{P}_{\Lambda_k} = \mathbbm{P}_{\Lambda_\ell}^{|\Lambda_k}$ for all $\Lambda_k \subseteq \Lambda_\ell \subseteq \Lambda$.  Intuitively, this means that $\mathbbm{P}_{\Lambda_\ell}$, the descriptor of the stochastic process on the times $\Lambda_\ell$, contains all information about subprocesses on fewer times. 

While an underlying process leads to a family of compatible finite probability distributions, the KET shows that the converse is also true. Any family of consistent probability distributions implies the existence of an underlying process. Specifically, the Kolmogorov extension theorem~\cite{kolmogorov_foundations_1956,feller_introduction_1968, tao_introduction_2011,breuer_theory_2007} defines the minimal properties finite probability distributions have to satisfy in order for an underlying process to exist:

\begin{theorem*}{\rm {[}Kolmogorov{]}}
Let $\Lambda$ be a set of times. For each finite $\Lambda_k \subseteq \Lambda$, let $\Pprob_{\Lambda_k}$ be a (sufficiently regular) $k$-step joint probability distribution. There exists an underlying stochastic process $\Pprob_\Lambda$ that satisfies $\Pprob_{\Lambda_k} = \Pprob_{\Lambda}^{|\Lambda_k}$ for all finite $\Lambda_k\subseteq \Lambda$ iff $\Pprob_{\Lambda_k} = \Pprob_{\Lambda_\ell}^{|\Lambda_k}$ for all $\Lambda_k\subseteq \Lambda_\ell \subseteq \Lambda$.
\end{theorem*}

In other words, if a family of joint probability distributions on finite sets of times satisfies a consistency condition there is an underlying stochastic process on $\Lambda$ that has the distributions $\left\{\Pprob_{\Lambda_k}\right\}_{\Lambda_k\subset \Lambda}$ as marginals. More precisely, the Kolmogorov extension theorem guarantees the existence of a probability \textit{measure} on an infinite product of measurable spaces if the respective measures on said spaces are compatible with each other in the above sense, and each of the measurable spaces is \textit{inner regular}~\cite{tao_introduction_2011}. That is, the measure of any set can be approximated by that of compact subsets. As we will consider our value spaces (\textit{i.e.}, the spaces of possible outcomes) to be $\mathbbm{R}$ or $\mathbbm{N}$ throughout this paper, the requirement of inner regularity of the considered probability distributions will always be automatically satisfied.\footnote{In particular, both $\mathbbm{R}$ and $\mathbbm{N}$ equipped with their standard topology are \textit{Borel spaces}. A probability measure on such a space is inner regular~\cite{tao_introduction_2011}.}

As stated above, the KET defines the notion of a classical stochastic process and reconciles the existence of an underlying process with its manifestations for finite times. It also enables the modelling of stochastic processes: Any mechanism that leads to finite joint probability distributions that satisfy a consistency condition is ensured to have an underlying process. For example, the proof of the existence of Brownian motion relies on the KET as a fundamental ingredient~\cite{wiener_differential_1966, Levy_1940, ciesielski_lectures_1966, bhattacharya_basic_2017}.

We emphasize that, in the (physically relevant) case where $\Lambda$ is an infinite set, the probability distribution $\mathbbm{P}_\Lambda$ is generally not experimentally accessible. For example, in the case of Brownian motion, the set $\Lambda$ could contain all times in the interval $[0,t]$ and each realization $\mathbf{i}_\Lambda$ would represent a possible continuous trajectory of a particle over this time interval. While we assume the existence of these underlying trajectories (and hence the existence of $\mathbbm{P}_\Lambda$) in experiments concerning Brownian motion, we often only access their finite time manifestations, \textit{i.e.}, $\mathbbm{P}_{\Lambda_k}$ for some $\Lambda_k$. The KET bridges the gap between the finite experimental reality and the underlying infinite stochastic process.

Loosely speaking, the KET holds for classical stochastic processes, because there is no difference between `doing nothing' and conducting a measurement but `not looking at the outcomes' (\textit{i.e.}, summing over the outcomes at a time). Put differently, the validity of the KET is based on the fundamental assumption that the interrogation of a system does not, on average, influence its state. Consequently, marginalization is the correct way to obtain the descriptor for fewer times and any classical stochastic process leads to compatible finite joint probability distributions; this compatibility is independent of whether the system was interrogated or not, and the converse also holds. This fails to be true in causal modelling scenarios.

\subsection{The KET and causal modelling}
The compatibility of joint probability distributions for different sets of times hinges on the fact that observations in classical physics do not alter the state of the system that is being observed. In contrast to passive interrogations, that merely reveal information, active interventions, like they are used in the case of causal modelling, on average change the state of the interrogated system. Thus the future statistics after an intervention crucially depends on how the system was manipulated and the prerequisite of compatible joint probability distributions is generally not fulfilled anymore. 

Consider, for example, the case of a pharmacologist that tries to understand the effect of different drugs they developed on a disease.
\begin{figure}[t!]
\centering
\subfigure[Without interventions.]
{
\includegraphics[width=0.95\linewidth]{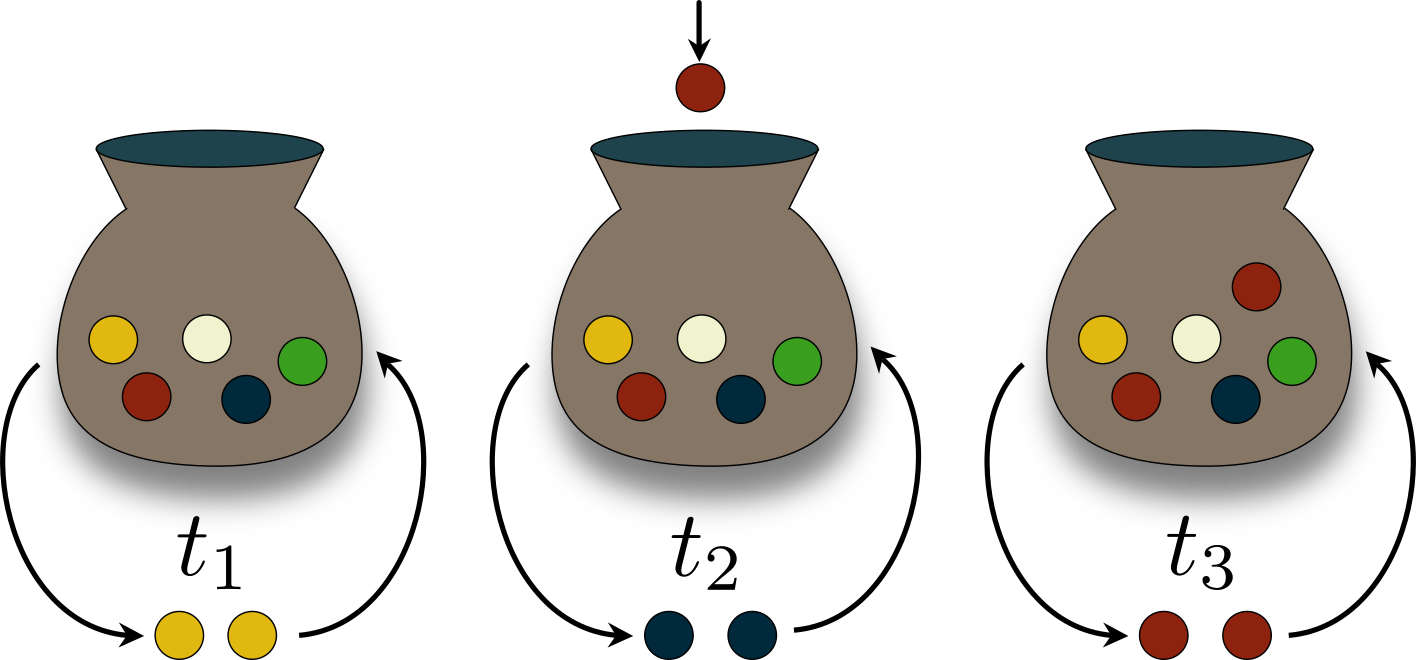}
}\\
\subfigure[With interventions.]
{
\includegraphics[width=0.95\linewidth]{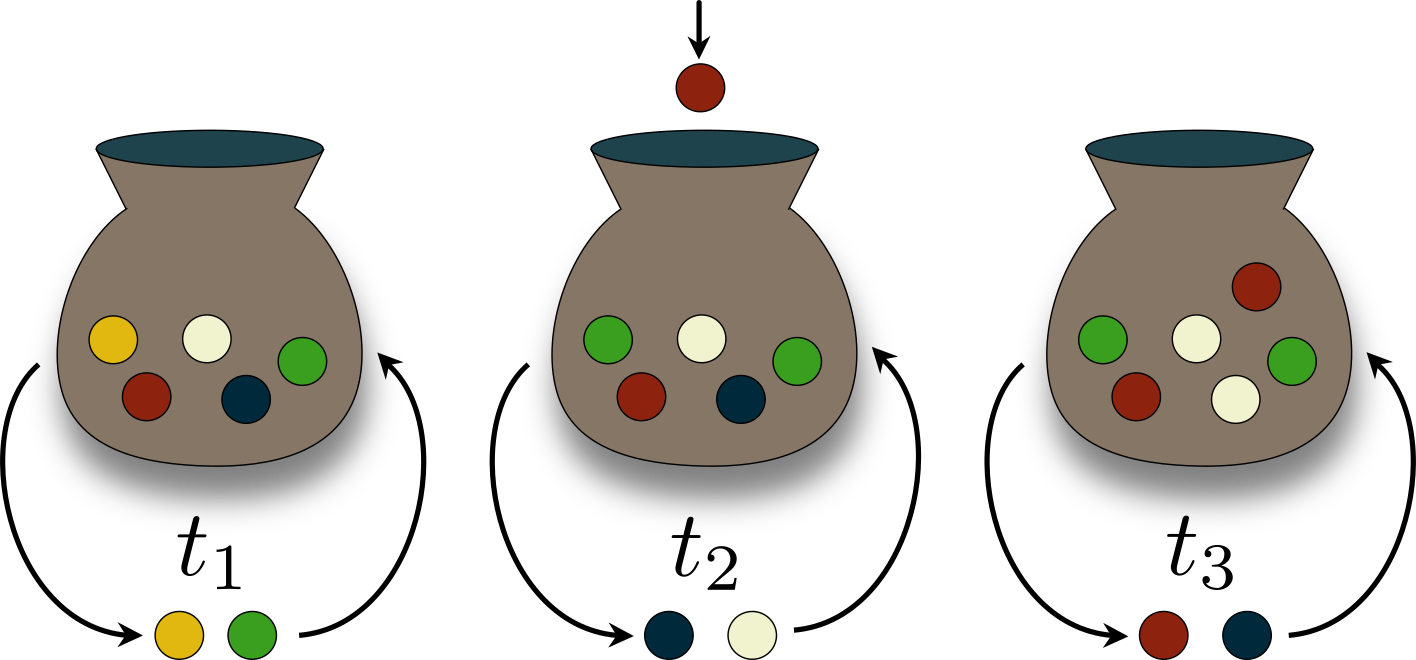}
}
\caption{\textbf{(a)} A possible three step process without intervention is the drawing with replacement of colored balls from an urn. In our example, independent of the actions of the experimenter, a red ball drops into the urn at $t_2$ (this could, \textit{e.g.}, represent the interaction with an uncontrollable environment.) The experimenter can deduce the joint probability distribution $\Pprob_{\{t_3,t_2,t_1\}}(c_3,c_2,c_1)$, to draw different sequences of colors. $\Pprob_{\{t_3,t_2,t_1\}}$ contains all distributions for fewer times, for example $\Pprob_{\{t_3,t_1\}}, \Pprob_{\{t_3,t_2\}}$, and $\Pprob_{\{t_1\}}$. \textbf{(b)} Instead of putting the same ball back in the urn, the experimenter could exchange it with a different color (for example, upon drawing yellow, they could replace it with green at $t_1$, replace blue with white at $t_2$ and replace red with blue at $t_3$). The respective replacement rules are encapsulated in the instruments $\Jcal_3$, $\Jcal_2$, and $\Jcal_1$. Now, from the probability distribution $\Pprob_{\{t_3,t_2,t_1\}}(c_{3},c_2,c_1|\Jcal_3,\Jcal_2,\Jcal_1)$, it is generally not possible to deduce probability distributions for fewer steps, like, \textit{e.g.}, $\Pprob_{\{t_3,t_1\}}(c_3,c_1|\Jcal_3,\Jcal_1)$, or $\Pprob_{\{t_3,t_2\}}(c_3,c_2|\Jcal_3,\Jcal_2)$. This lack of consistency can not be remedied by simple relabeling of the times due to the red ball that drops into the urn at $t_2$. Note that if all instruments are the idle instrument, the case with interventions coincides with the case without interventions.}
\label{fig::urne}
\end{figure}
In our simplified example, let the disease have two different symptoms $S_a$ and $S_b$, and denote the absence of symptoms $S_c$. Whenever the pharmacologist observes $S_a$, they administer drug $D_a$, whenever they observe symptom $S_b$ they administer drug $D_b$, and whenever they observe $S_c$ they do nothing; this choice of actions defines an instrument $\mathcal{J}$. Running their trial with sufficiently many patients, the pharmacologist can deduce probability distributions for the occurrence of symptoms over time, given the drugs that were administered. For example, if the drugs were administered on three consecutive days, they would have obtained a probability distribution of the form $\mathbbm{P}_{\Lambda_3}(s_{3},s_{2},s_{1}|\Jcal_3=\Jcal,\Jcal_2=\Jcal,\Jcal_1=\Jcal)$, where $s_{\alpha} \in \{S_a,S_b,S_c\}$, and the instruments (\textit{i.e.}, the drug administration rule) are the same each day. However, this data would not allow them to find out what would have happened, had they \textit{not} administered drugs on day two, \textit{i.e.}, $\sum_{s_2} \mathbbm{P}_{\Lambda_3}(s_{3},s_{2},s_{1}|\Jcal_3,\Jcal_2,\Jcal_1) \neq \mathbbm{P}_{\Lambda_3\setminus \{t_2\}}(s_{3},s_{1}|\Jcal_3,\Jcal_1)$; intermediate interventions change the state of the interrogated system, and hence the future statistics that are being observed; for another, more numerically tangible example, see Fig.~\ref{fig::urne}. Consequently, probability distributions do generally not satisfy compatibility conditions when interventions are allowed.

Compatibility can fail to hold whenever the system of interest is actively interrogated. In particular, it fails to hold in quantum mechanics, where even projective measurements in general change the state of a system on average, and interventions are not just an experimental choice but unavoidable.

\section{(Quantum) Stochastic processes with interventions}
\label{sec::QuantStoch}
\subsection{The KET in QM}
\label{subsec::KET_QM}
As hinted at throughout this work, descriptions of quantum mechanical processes must necessarily account for the fundamental invasiveness of measurements, which renders the KET invalid for the same reason that some choices of intervention do in the case of classical causal modelling. To see how even projective measurements in quantum mechanics lead to families of probability distributions that do not satisfy the KET, consider the following concatenated Stern-Gerlach experiment: Let the initial state of a  spin-$\frac{1}{2}$ particle be $\ket{+} = \tfrac{1}{\sqrt{2}}(\ket{\uparrow} + \ket{\downarrow})$, where $\ket{\uparrow}$ and $\ket{\downarrow}$ are the spin-up and spin-down state in the $z$-direction, respectively. Now, we measure the state sequentially in the $z$-, $x$- and $z$-directions at times $t_1, t_2$ and $t_3$. These measurements have the possible outcomes $\{\uparrow,\downarrow\}$ and $\{\rightarrow,\leftarrow\}$ for the measurement in $z$- and $x$-direction, respectively. It is easy to see that the probability for any possible sequence of outcomes is equal to $1/8$. For example, we have
\begin{align}
    \notag
    &\Pprob_{\Lambda_3}(\uparrow,\rightarrow,\uparrow|\Jcal_z,\Jcal_x,\Jcal_z) \\
    \label{eqn::ProbSternGerlach} &\phantom{asdfasda}= \Pprob_{\Lambda_3}(\uparrow,\leftarrow,\uparrow|\Jcal_z,\Jcal_x,\Jcal_z) = \frac{1}{8}\, ,
\end{align}
where $\Jcal_z$ and $\Jcal_x$ represent the instruments used to measure in the $z$- and $x$- direction respectively, and $\Lambda_3 = \{t_3,t_2,t_1\}$. Summing over the outcomes at time $t_2$, we obtain the marginal probability $\Pprob_{\Lambda_3}^{|\{t_1,t_3\}}(\uparrow,\uparrow|\Jcal_z,\Jcal_z) = 1/4$. However, by considering the case where the measurement is not made at $t_2$, it is easy to see that $\Pprob_{\{t_3,t_1\}}(\uparrow,\uparrow|\Jcal_z,\Jcal_z) = 1/2$. The intermediate measurement changes the state of the system, and the corresponding probability distributions for different sets of times are not compatible anymore~\cite{BreuerEA2016, shrapnel_causation_2018}. 

It is important to highlight the close relation of this breakdown of consistency and the violation of Leggett-Garg inequalities in quantum mechanics~\cite{leggett_quantum_1985,emary_leggettgarg_2014}. The assumption of consistency between descriptors for different sets of times that is crucial for the derivation of the KET subsumes the assumptions of realism \textit{per se} and noninvasive measurability that are the basic principles leading to the derivation of Leggett-Garg inequalities: While realism \textit{per se} implies that joint probability distributions for a set of times can be expressed as marginals of a joint probability distribution for more times, non-invasiveness means that all finite distributions are marginals of the \textit{same} distribution. For example, the two-step joint probability distributions $\mathbbm{P}_{\{t_2,t_1\}}$, $\mathbbm{P}_{\{t_3,t_2\}}$, and $\mathbbm{P}_{\{t_3,t_1\}}$, that are considered in the Leggett-Garg scenario are all marginals of a three-step distribution $\mathbbm{P}_{\{t_3,t_2,t_1\}}$. As soon as non-invasiveness and/or realism \textit{per se} do not hold, the KET can fail and Legget-Garg inequalities can be violated.

Nevertheless, there should be some compatibility between descriptors for different sets of times; the breakdown of the KET should be a problem of the formalism rather than a physical fact. We now show that a change of perspective enables one to prove an extension theorem in quantum mechanics and any theory with interventions. 

\subsection{Instruments and Combs}
\label{subsec::GET}
Processes involving interventions, including quantum processes and those in classical causal modelling, do not lead to compatible joint probability distributions for different sets of times in general. This problem can be remedied by assuming the standpoint of quantum causal modelling, and choosing a description of such stochastic processes that takes interventions and their corresponding change of the system into account. With this description, it is possible to recover a compatibility property that is satisfied by any process with interventions, and a generalized extension theorem can be derived. 

As in the classical causal modelling case, in quantum mechanics, an experimenter can choose an instrument $\Jcal_\alpha$ at each time $t_\alpha$, and every outcome $i_\alpha$ corresponds to a particular transformation of the system that is interrogated. Denoting the Hilbert space of the system at $t_\alpha$ by $\Hcal_\alpha$, mathematically, an observation of outcome $i_\alpha$ given the instrument $\Jcal_\alpha$ corresponds to a (trace non-increasing) completely positive (CP) map $\Mcal_{i_\alpha}:\Bcal_1(\Hcal_\alpha^\inp) \rightarrow \Bcal_1(\Hcal_\alpha^\out)$ that describes the change of the state of the system~\cite{Lindblad1979,davis_quantum_1976}. Here, $\Bcal_1(\Hcal_\alpha)$ denotes the space of trace class operators on $\Hcal_\alpha$ -- which in the finite dimensional case coincides with the set of bounded operators $\Bcal(\Hcal_\alpha)$ on $\Hcal_\alpha$ -- and we account for the possibility that $\Mcal_{i_\alpha}$ creates or discards degrees of freedom by distinguishing between its input ($\inp$) and output ($\out$) spaces. The set of possible CP maps an instrument comprises add up to a completely positive trace preserving (CPTP) map $\Mcal_\alpha = \sum_{i_\alpha} \Mcal_{i_\alpha}$, which describes the overall average transformation applied by the instrument. While the CP maps the instrument comprises can only be implemented probabilistically, the corresponding overall CPTP map can be performed deterministically, \textit{i.e.}, with unit probability. In contrast to classical physics without interventions, where the introduction of maps (or \textit{events}  more generally~\cite{shrapnel_causation_2018}) is superfluous, it is fundamentally unavoidable in quantum mechanics, as well as in more general probabilistic theories~\cite{chiribella_probabilistic_2010,shrapnel_causation_2018}; every measurement alters the state of the system of interest, and a full description of a temporal process necessitates knowledge of how the system is changed at each time.

As in the example from Sec.~\ref{subsec::KET_QM}, an experimenter could choose to measure a system in different bases. The projective measurement in the basis $\{\ket{i}\}$ at $t_\alpha$ of a state $\rho$ that yields outcome $i_{\alpha}$ would be described by a CP map of the form $\Mcal_{i_\alpha}[\rho] = \bra{i_{\alpha}}\rho\ket{i_{\alpha}} \ketbra{i_{\alpha}}{i_{\alpha}}$, where $\ket{i_\alpha }\in \{\ket{i}\}$. More generally, the measuring instrument need not preserve the measured state of the system, but could replace it entirely; upon measuring an outcome $i_\alpha$ (corresponding to a projection on a state $\ket{i_\alpha}$), a different instrument could leave the system in the state $\rho_{i_\alpha}$, with a resulting CP map $\Mcal_{i_\alpha}[\rho] = \bra{i_\alpha}\rho\ket{i_\alpha} \rho_{i_\alpha}$. In the most general case, the experimenter -- at a time $t_\alpha$ -- could perform \textit{any} (trace non-increasing) CP map, including deterministic operations such as unitary transformations. We will employ the convention that, for a given instrument $\Jcal_\alpha$, the CP map corresponding to the outcome $i_\alpha$ is denoted by $\Mcal_{i_\alpha}$, and we will denote the complex vector space that is spanned by all of these CP maps by $\Lcal_\alpha$.

In this language, each realization of an experiment corresponds to a (possibly temporally correlated, see below) sequence of CP maps that transform the system at a series of times. The set of possible CP maps that could be applied is dictated by the choice of instruments used to interrogate the system in question. A quantum process is fully characterized once \textit{all} of the probabilities $\Pprob_{\Lambda_k}(i_k,\dots,i_1|\Jcal_k,\dots, \Jcal_1)$ for each such sequence with all possible instruments are known. Having all of these probability distributions at hand allows one to deduce the causal structure of a process, \textit{i.e.}, it is the basis of quantum causal modelling~\cite{1367-2630-18-6-063032}.

\begin{figure}
\centering
\includegraphics[scale=0.4]{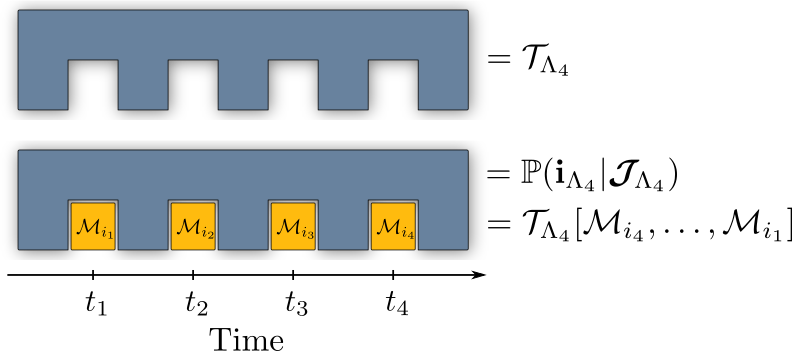}
\caption{\textit{Graphical representation of a four step quantum comb.} A four step comb can be represented as an object with four slots (each slot corresponds to a time $t_\alpha \in \Lambda_4$); it encodes all multi-time correlations between observables at those times. The probability to observe the outcomes $i_4,\dots,i_1$ given the instruments $\Jcal_4,\dots,\Jcal_1$ is obtained by inserting the corresponding CP maps into the comb, \textit{i.e.}, by letting $\mathcal{T}_{\Lambda_4}$ act on $\Mcal_{i_4},\dots, \Mcal_{i_1}$. }
\label{fig::Def_Proc}
\end{figure}

Written more succinctly, a $k$-step quantum process is fully characterized by an object $\Tcal_{\Lambda_k}$ that maps sequences of CP maps to probabilities. Specifically, this means that $\Tcal_{\Lambda_k}$ is a CP map $\Tcal_{\Lambda_k}: \Lcal_{\Lambda_k} \rightarrow \mathbbm{C}$, where $\Lcal_{\Lambda_k} = \bigotimes_{\alpha \in \Lambda_k} \Lcal_{\alpha}$, such that $\Tcal_{\Lambda_k}[\Mcal_{i_k},\dots,\Mcal_{i_1}]$ yields the probability $\Pprob_{\Lambda_k}(i_k,\dots,i_1|\Jcal_k,\dots,\Jcal_1)$ to obtain the outcomes $i_k,\dots, i_1$ given the choices of instruments $\Jcal_k,\dots,\Jcal_1$ (see Fig.~\ref{fig::Def_Proc}). In this sense, $\Tcal_{\Lambda_k}$ represents a Born rule for temporal processes~\cite{chiribella_memory_2008,shrapnel_updating_2017}. The mapping $\Tcal_{\Lambda_k}$ is a completely positive multilinear functional that can be reconstructed in a finite number of experiments~\cite{modi_operational_2012, pollock_non-markovian_2018, milz_reconstructing_2016, milz_introduction_2017}. 

For example, a $k$-step quantum process could be of the form 
\begin{align}
\notag
    &\Tcal_{\Lambda_k}[\Mcal_{i_k},\dots,\Mcal_{i_1}] \\
    \notag
    &= \tr\{\Mcal_{i_k} \circ \Ecal_{k-1} \circ \Mcal_{i_{k-1}} \circ \Ecal_{k-2} \cdots \\
    &\phantom{asdfggggggggasd}\cdots \circ \Mcal_{i_2} \circ \Ecal_1 \circ \Mcal_1[\rho]\}\,,
\end{align}
where $\{\Ecal_i\}$ are CPTP maps and $\rho$ is a fixed state of the system of interest. More generally, $\Tcal_{\Lambda_k}$ can describe a process with memory -- \textit{i.e.}, a non-Markovian process~\cite{pollock_non-markovian_2018} -- in which case it would be of the form 
\begin{align}
\label{eqn::non_Markov}
\notag
    &\Tcal_{\Lambda_k}[\Mcal_{i_k},\dots,\Mcal_{i_1}] \\
    \notag &= \tr\{(\Mcal_{i_k} \otimes \Ical_e) \circ \Ucal_{k-1} \circ (\Mcal_{i_{k-1}} \otimes \Ical_e) \circ \cdots \\
    &\phantom{=}\cdots \circ \Ucal_1 \circ (\Mcal_{i_1} \otimes \Ical_e)[\rho_{se}]\} ,
\end{align}
where $\rho_{se}$ is a (possibly correlated) state on the system of interest and an additional environment, and the maps $\{\Ucal_i\}$ can be chosen to be unitary evolutions on the system-environment space. As information from the past can be propagated through the additional environment, processes that are described via Eq.~\eqref{eqn::non_Markov} generally display non-trivial memory effects~\cite{pollock_non-markovian_2018}. 

If $\Lambda_k$ corresponds to an ordered set of times, then \textit{every} $k$-step process on $\Lambda_k$ can be written in the form of Eq.~\eqref{eqn::non_Markov}~\cite{chiribella_theoretical_2009, pollock_non-markovian_2018}. On the other hand, if the process at hand were to not abide by a clear causal order, then it would not possess a representation of this form, but could still display non-trivial correlations between different events~\cite{PhysRevA.88.022318,OreshkovETAL2012}. The respective causal ordering that $\Tcal_{\Lambda_k}$ is compatible with imposes further requirements on its properties; besides being CP, $\Tcal_{\Lambda_k}$ has to be properly normalized. Naturally, $\Tcal_{\Lambda_k}$ has to yield unit probability when acting on an operation that can be implemented deterministically. Consequently, we have 
\begin{gather}
    \label{eqn::unitProb}
    \Tcal_{\Lambda_k}[\Mcal_{k},\dots,\Mcal_{1}] = 1
\end{gather} 
for all sequences $\Mcal_{k},\dots,\Mcal_{1}$ of CPTP maps. Additionally, depending on the underlying causal structure, the respective deterministic operations an experimenter performs could be correlated -- both in a classical and a quantum way. For example, if $\Lambda_k$ is an ordered set of times, the choice of instrument used to interrogate the system at a time $t_{\alpha'}\in \Lambda_k$ could be conditioned on the outcomes of all measurements at times $\Lambda_k \ni t_\alpha < t_{\alpha'}$. Put more generally, in quantum mechanics, deterministic operations are \textit{all} operations that can be realized by preparing ancillary states, performing unitary operations, and discarding degrees of freedom in a temporally ordered fashion that agrees with the ordering given by $\Lambda_k$.

On the other hand, in situations where $\Lambda_k$ does not correspond to a temporally ordered set, or, more generally, a partially ordered set corresponding to a definite causal order~\cite{chiribella_theoretical_2009}, but rather a set of labels for different laboratories with an unclear causal ordering, then such non-trivially correlated operations are not considered to be experimentally implementable, and the corresponding linear functionals $\mathcal{T}_{\Lambda_k}$ only have to satisfy Eq.~\eqref{eqn::unitProb} (and be CP), with no additional restrictions~\cite{OreshkovETAL2012}.

Finally, in operational probabilistic theories (OPTs), the set of deterministically implementable operations is the set of operations obtained from concatenating deterministic \textit{preparations}, deterministic \textit{transformations}, and deterministic \textit{effects} in a well-defined order that agrees with the causal structure of the respective OPT~\cite{dariano_quantum_2017}.

In anticipation of later proofs, it is worth briefly discussing the set of operations that proper physical linear functionals on $k$ times can be meaningfully applied to. As mentioned above, in quantum mechanics, a deterministically implementable operation is one that can be decomposed as a causally ordered concatenation of state preparations, unitary operations and a final discarding of degrees of freedom. Analogously, a general (temporally correlated) quantum measurement that an experimenter performs can always be decomposed as a causally ordered concatenation of state preparations, unitary operations and a final projective measurement on some degrees of freedom~\cite{chiribella_theoretical_2009}.

More concretely, setting $\Hcal_{\Lambda_k} = \bigotimes_{\alpha \in \Lambda_k} \Hcal_\alpha$, such scenarios would be described by a collection of temporally non-local (trace non-increasing) CP maps $\Mcal^{\Lambda_k}_{\gamma}: \Bcal_1(\Hcal^\inp_{\Lambda_k}) \rightarrow \Bcal_1(\Hcal^\out_{\Lambda_k})$, where each of the maps $\Mcal^{\Lambda_k}_{\gamma}$ corresponds to a possible measurement outcome. Overall, the respective probabilities have to add to unity, which implies that $\Mcal_{\Lambda_k} = \sum_\gamma \Mcal^{\Lambda_k}_{\gamma}$ corresponds to a deterministic operation. Such a collection $\{\Mcal^{\Lambda_k}_{\gamma}\}$ of maps is known as  a \textit{generalized instrument} or \textit{tester} in the literature~\cite{chiribella_theoretical_2009, chiribella_memory_2008, chiribella_optimal_2016} (see~\cite{chiribella_theoretical_2009} for a thorough discussion of their properties). 

Naturally, a physically reasonable mapping $\Tcal_{\Lambda_k}$ for such a causal scenario has to satisfy $0\leq\Tcal_{\Lambda_k}[\Mcal^{\Lambda_k}_{\gamma}]\leq 1$ for all possible tester elements on $\Lambda_k$ and  $\Tcal_{\Lambda_k}[\Mcal_{\Lambda_k}] = 1$ for any temporally correlated operation that can be implemented deterministically. Likewise, in scenarios where no \textit{a priori} causal order is assumed, $\Tcal_{\Lambda_k}$ only has to satisfy Eq.~\eqref{eqn::unitProb} as well as $0 \leq \Tcal_{\Lambda_k}[\Mcal_{i_1},\dots,\Mcal_{i_k}]\leq 1$ for any collection $\{\Mcal_{i_k}\}$ of CP maps. Finally, in OPTs, `tester elements' are given by all operations that can be realized by concatenating deterministic preparations, deterministic transformations, and \textit{probabilisitc} effects in an order that agrees with $\Lambda_k$. As before, a physically reasonable mapping $\Tcal_{\Lambda_k}$ would have to yield unit probability on deterministically implementable operations, and a probability $p\in [0,1]$ for any operation that cannot be implemented deterministically.

In each case, the set of operations that a physically reasonable functional $\Tcal_{\Lambda_k}$ can be applied to forms a convex set: if $\{\Mcal^{\Lambda_k}_{\gamma}\}$ is a set of (probabilistically and deterministically) implementable operations, then 
\begin{gather}
\label{eqn::convex}
    \Tcal_{\Lambda_k}\left[\sum_\gamma \mu_\gamma \Mcal^{\Lambda_k}_{\gamma}\right] \in [0,1]\, .
\end{gather} 
where $\mu_\gamma \geq 0$ and $\sum_\gamma \mu_\gamma \leq 1$. We will denote this set of operations on times $\Lambda_k$ that $\Tcal_{\Lambda_k}$ can meaningfully act on $\Kcal_{\Lambda_k}$. 

Since the respective normalization of the families of linear functionals $\Tcal_{\Lambda_k}$ that we consider, as well as the particular underlying theory is not of importance for the proof and/or the applicability of the main theorem of the paper, we will assume an agnostic standpoint and adhere to the following convention:

Throughout the remainder of this Paper, a multilinear functional $\Tcal_{\Lambda_k}: \Lcal_{\Lambda_k} \rightarrow \mathbbm{C}$ that is positive on $\Kcal_{\Lambda_k}$, and which yields unit probability on \textit{all} deterministically implementable operations that the underlying causal structure (and underlying theory) allows, will be referred to as a $k$-step \textit{comb}, following Refs.~\cite{chiribella_transforming_2008, chiribella_quantum_2008, chiribella_theoretical_2009}. For the particular case of quantum mechanics, we would additionally demand that the comb is CP. 

Note that, in the convention above, $\Kcal_{\Lambda_k}$ and its associated set of $k$-step combs depend on the respective scenario one considers. Additionally, as $\Tcal_{\Lambda_k}$ is positive on all elements of $\Kcal_{\Lambda_k}$, we automatically have $\Tcal_{\Lambda_k}[\Mcal_{\gamma}^{\Lambda_k}] \leq 1$, since for any $\Mcal_{\gamma}^{\Lambda_k}\in \Kcal_{\Lambda_k}$, there exists a $\Mcal_{\gamma'}^{\Lambda_k}\in \Kcal_{\Lambda_k}$ such that $\Mcal_{\gamma}^{\Lambda_k} + \Mcal_{\gamma'}^{\Lambda_k}\in \Kcal_{\Lambda_k}$ is a deterministically implementable operation.
Importantly, outside the set $\Kcal_{\Lambda_k}$, $\Tcal_{\Lambda_k}$ can yield `probabilities' that exceed $1$, even when acting on maps that may be deterministically implementable in other scenarios. This is due to the fact that the action of a comb $\Tcal_{\Lambda_k}$ on a deterministic map that is not causally ordered in a compatible way can lead to causal loops, and thus non-sensical results; for example, letting a comb that is ordered $t_1$ before $t_2$ act on an operation that is ordered $t_2$ before $t_1$ will lead to causal loops and is, as such, not meaningful. 

In what follows, we will predominantly phrase our statements with respect to CP maps (\textit{i.e.}, for the case of quantum mechanics), with the understanding that generalization to other theories is always straightforwardly possible.

A comb $\Tcal_{\Lambda_k}$ contains all the multi-time correlations necessary to fully characterize a $k$-step quantum process. While the CP maps $\Mcal_{i_\alpha}$ change the state of the system, they do not change the $k$-step process given by $\Tcal_{\Lambda_k}$. Loosely speaking, the comb contains all parts of the dynamics that are not manipulated by the experimenter. This is analogous to the way in which the preparation of an initial state and the measurement of the final state in quantum process tomography do not influence the underlying dynamics (\textit{i.e.}, the CPTP map connecting input and output state).  

Just as in the classical case, the knowledge of all relevant joint probability distributions (\textit{i.e.}, the knowledge of $\Tcal_{\Lambda_k}$) allows one to deduce causal relations between the $k$ events in $\Lambda_k$. We emphasize that classical causal modelling is included in this quantum causal modelling framework as a special case. Whenever a system is measured and prepared in a fixed basis (using a classical instrument), and the process $\Tcal_{\Lambda_k}$ also preserves this basis, the result is a set of joint distributions consistent with a classical causal model. From Observation~\ref{lem::ccm_csp}, it also follows that classical stochastic processes without interventions can be described by the same framework. 

\subsection{Generalized extension theorem (GET)}
\begin{figure}
\centering
\includegraphics[scale=0.61]{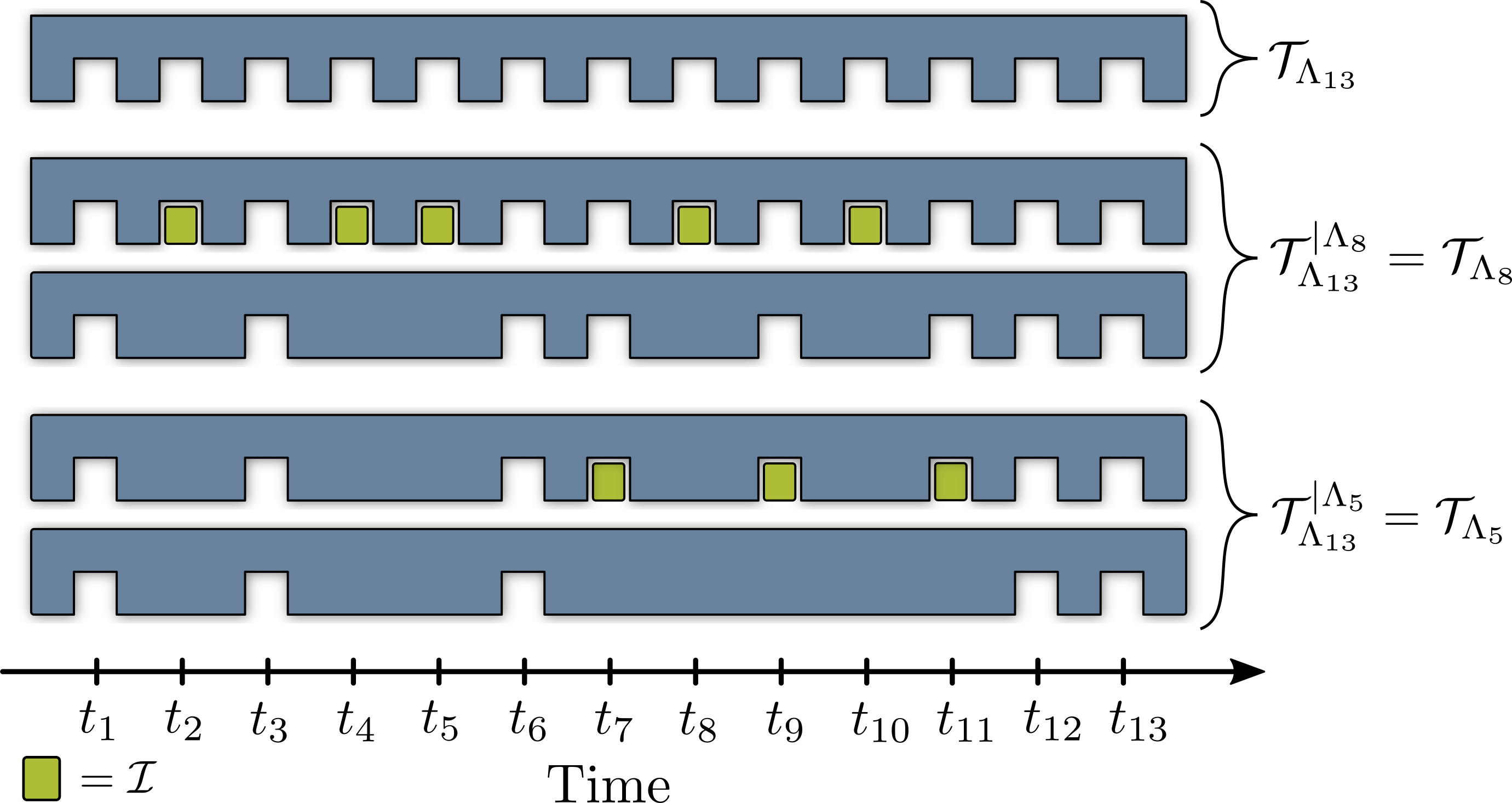}
\caption{\textit{Consistency condition for combs.} If there is an underlying process, any comb $\mathcal{T}_{\Lambda_k}$ can be obtained from $\mathcal{T}_{\Lambda_\ell
}$ by letting $\mathcal{T}_{\Lambda_\ell}$ act on the identity map at the excessive times. Here, for the sets of times $\Lambda_{13} = \{t_{13},\dots,t_1\}$, $\Lambda_8 = \{t_{13},t_{12},t_{11},t_9,t_7,t_6,t_3,t_1\}$ and $\Lambda_5 = \{t_{13},t_{12},t_6,t_3,t_1\}$ we depict the containment of the comb $\mathcal{T}_{\Lambda_8}$ in $\mathcal{T}_{\Lambda_{13}}$ and the containment of $\mathcal{T}_{\Lambda_5}$ in both $\Tcal_{\Lambda_{13}}$ and $\Tcal_{\Lambda_8}$.}
\label{fig::Containment}
\end{figure}

With the complete description on finite sets of times at hand, we can determine the compatibility condition between related combs. A family of combs that stems from an underlying (open) dynamics fulfills a natural consistency condition~\cite{pollock_non-markovian_2018}; for any two sets of times $\Lambda_{k} \subseteq \Lambda_\ell$, the comb $\Tcal_{\Lambda_k}$ can be obtained from $\Tcal_{\Lambda_\ell}$ by letting it act on identity operations $\Ical_{t_\alpha}$ (with $\Ical_{t_\alpha}[\rho] = \rho$ for any state of the system $\rho$ at time $t_\alpha$) at times $t_\alpha \in \Lambda_\ell\setminus \Lambda_k$, \textit{i.e.},
\begin{gather}
\label{eqn::Consistency}
\Tcal_{\Lambda_{k}}[\,\cdot\,] = \Tcal_{\Lambda_\ell} \left[\bigotimes_{t_\alpha \in \Lambda_\ell\setminus \Lambda_k}\Ical_{t_\alpha},\,\cdot\,\right] := \Tcal_{\Lambda_\ell}^{|\Lambda_k}[\,\cdot\,]\, ,
\end{gather}
where we have employed the shorthand notation $\bigotimes_{t_\alpha \in \Lambda_\ell\setminus \Lambda_k}\Ical_{t_\alpha}$ to signify that the identity operation was `implemented' at each time $t_\alpha \in \Lambda_\ell\setminus \Lambda_k$. The graphical representation of Eq.~\eqref{eqn::Consistency} is depicted in Fig.~\ref{fig::Containment}.

It is important to note the difference between Eq.~\eqref{eqn::Consistency} and the consistency condition for classical stochastic processes, stemming from the stronger notion of `doing nothing' in the quantum case. If there is an underlying process, any comb can be obtained from one that applies to a larger set of times by letting it act on the identity  map, which leaves any state unchanged, at the excessive times. This is by no means the same as computing the marginals of families of probability distributions that have been obtained for a fixed set of measurement instruments, which will only preserve states which are diagonal in a fixed basis. We recover descriptors for different sets of times that are compatible with each other only when we switch to a causal modelling description of the process. From this, we obtain our main result, the generalized extension theorem (GET): 
\begin{theorem*}[Generalized extension theorem]
\label{thm::Quolmogorov}
Let $\Lambda$ be a set of times. For each finite $\Lambda_k \subseteq \Lambda$ let $\Tcal_{\Lambda_k}$ be a $k$-step comb. There exists a general stochastic process $\Tcal_\Lambda$, \textit{i.e.}, a multilinear CP functional defined on all times in $\Lambda$, that satisfies $\Tcal_{\Lambda_k} = \Tcal_{\Lambda}^{|\Lambda_k}$, as defined in Eq.~\eqref{eqn::Consistency}, for all finite $\Lambda_k\subseteq \Lambda$ iff $\Tcal_{\Lambda_k} = \Tcal_{\Lambda_\ell}^{|\Lambda_k}$ for all $\Lambda_k\subseteq \Lambda_\ell\subseteq \Lambda$.
\end{theorem*}

Note that we make no assumption about the cardinality of the number of outcomes at each time $t_\alpha$. The GET holds independently of whether there are finitely, countably or uncountably many possible outcomes.

Importantly, the GET is qualitatively distinct from {quantum marginal problems}~\cite{tyc_quantum_2015}. The question it answers is \textit{not} whether, for a given collection of combs $\Tcal_{\Lambda_k}, \Tcal_{\Lambda_\ell}, \dots$, there exists a comb $\Tcal_{\Lambda_k\cup\Lambda_\ell\cup \cdots}$ that has all of them as marginals. Rather, it starts from the assumption that there exists a family of combs $\{\Tcal_{\Lambda_k}\}$ defined on \textit{all} finite subsets of $\Lambda$ and shows that this family can be extended to a comb $\Tcal_\Lambda$ if the family $\{\Tcal_{\Lambda_k}\}$ satisfies the consistency conditions laid out in the above theorem. For the case that $\Lambda$ is a finite set, the theorem is thus trivial, as the desired comb is simply the comb of the family $\{\Tcal_{\Lambda_k}\}$ that is defined on the largest set of times (that is, on $\Lambda$ itself). The importance of the GET lies in the case where $\Lambda$ is an infinite set. There, it shows that a family of finite combs satisfying proper consistency conditions suffices to define an underlying comb on all times in $\Lambda$.

The proof of the GET proceeds analogously to that of the original Kolmogorov extension theorem, presented, \textit{e.g.}, in Ref.~\cite{tao_introduction_2011}. It can be broken up into two main parts: \textbf{(i)} The consistency property is used to define a unique comb $\Tcal^\sharp_{\Lambda}$ on a `sufficiently large' container space $\Lcal_{\Lambda}^\sharp$. \textbf{(ii)} It is shown that $\Tcal^\sharp_\Lambda$ is linear and bounded on a subset $\Kcal^\sharp_{\Lambda}$ of said container space and can thus be extended to a linear functional $\overline{\Tcal}^\sharp_{\Lambda}$ fulfilling the properties of the desired comb $\Tcal_\Lambda$ on the closure of $\Kcal^\sharp_{\Lambda}$. As in the classical case~\cite{breuer_theory_2007}, the underlying stochastic process characterized by $\Tcal_{\Lambda}$ is -- unlike $\overline{\Tcal}^\sharp_{\Lambda}$ -- not necessarily unique. Since the action of all possible $\Tcal_{\Lambda}$ coincides with the unique $\overline{\Tcal}^\sharp_{\Lambda}$ on the correct set of operations, and hence yields the correct finite combs $\Tcal_{\Lambda_k}$, this non-uniqueness cannot be detected experimentally and does not constitute a practical problem.

\begin{proof} To begin with, a short comment on the spaces that the respective objects we will deal with are defined on is in order. Firstly, for ease of notation, we assume all CP maps that the combs act on to have the same input and output space, \textit{i.e.}, they do not create or discard degrees of freedom.\footnote{A generalization to maps with distinct input and output spaces is straightforward.} Consequently, at each time $t_\alpha$, we have $\Mcal_{i_\alpha}: \Bcal_1(\Hcal_\alpha) \rightarrow \Bcal_1(\Hcal_\alpha)$. As before, for any finite $\Lambda_k$, we will employ the naming conventions $\Hcal_{\Lambda_k}$, $\Bcal_1(\Hcal_{\Lambda_k})$, and $\Lcal_{\Lambda_k}$ for the Hilbert space $\Hcal_{\Lambda_k} = \bigotimes_{\alpha \in \Lambda_k} \Hcal_\alpha$, the space of trace class operators thereon, and the vector space spanned by the CP maps on $\Bcal_1(\Hcal_{\Lambda_k})$, respectively.   Notably, in the finite dimensional case, $\Bcal_1(\Hcal_{\Lambda_k})$ coincides with $\Bcal(\Hcal_{\Lambda_k})$, the space of bounded operators on $\Hcal_{\Lambda_k}$, while in general we have $\Bcal_1(\Hcal_{\Lambda_k}) \subset \Bcal(\Hcal_{\Lambda_k})$. As the generalization to infinite dimensional Hilbert spaces $\Hcal_\alpha$ does not bring additional technical difficulties, we will not assume $\mathrm{dim}(\Hcal_\alpha)$ to be finite in what follows. 

Since we will deal with operations at \textit{each} time $t_\alpha \in \Lambda$, the natural Hilbert space to consider is the -- possibly uncountably infinite -- tensor product $\Hcal_\Lambda := \bigotimes_{\alpha \in \Lambda} \Hcal_\alpha$. Such infinite products of Hilbert spaces have been defined in Def. $3.5.1$ of~\cite{neumann_infinite_1939}. For their construction, let $\{f_\alpha|f_\alpha\in \Hcal_\alpha\}_{\alpha\in \Lambda}$ be a $C$-sequence, \textit{i.e.}, a sequence of elements of the Hilbert spaces $\Hcal_\alpha$ such that $\prod_{\alpha\in \Lambda} \|f_\alpha\|$ converges.\footnote{See~\cite{neumann_infinite_1939} for a detailed discussion of the convergence of uncountably infinite products and sums.} On any such $C$-sequence, one can define particular linear functionals 
\begin{gather}
\label{eqn::Phi0}
    \Phi^r(f_\alpha;\alpha \in \Lambda) := \prod_{\alpha \in \Lambda} (f_\alpha^r,f_\alpha)\, ,
\end{gather}
where $\{f^r_\alpha|f^r_\alpha\in \Hcal_\alpha\}_{\alpha\in \Lambda}$ is a $C$-sequence, and $(\cdot,\cdot)$ is the respective inner product of $\Hcal_\alpha$. In a slight abuse of notation, this functional can be written $\Phi^r = \bigotimes_{\alpha \in \Lambda} f_\alpha^r$. With this, we can define the space $\Fcal^0$ as the finite linear span of all such elements $\Phi^r$, \textit{i.e.}, 
\begin{gather}
    \Fcal^0 = \{\Theta|\Theta = \sum_{r=1}^N\Phi^r\}\, .
\end{gather} 
This space is equipped with a well-defined scalar product~\cite{neumann_infinite_1939}: Taking two elements $\Phi_1 = \sum_{r=1}^{N_1} \bigotimes_{\alpha \in \Lambda} f_{\alpha}^r$ and $\Phi_2 = \sum_{s=1}^{N_2} \bigotimes_{\alpha \in \Lambda} g_{\alpha}^s$ of $\mathcal{F}_0$, we can set
\begin{gather}
\label{eqn::scalar}
   (\Phi_1,\Phi_2):= \sum_{r=1}^{N_1}\sum_{s=1}^{N_2}\prod_{\alpha \in \Lambda} (f_\alpha^r,g_\alpha^s)\, .
\end{gather}
With this, we obtain the space $\Hcal_\Lambda$ as the set of all limits of Cauchy sequences in $\Fcal^0$. Specifically, all elements $\varphi \in \Hcal_\Lambda$ are such that there exists a Cauchy sequence $\{\Theta_s; \Theta_s \in \Fcal^0\}$ that converges to $\varphi$ in the sense that
\begin{gather}
    \varphi(f_\alpha;\alpha \in \Lambda) = \lim_{s\to \infty} \Theta_s(f_\alpha;\alpha \in \Lambda)\, ,
\end{gather}
for all $C$-sequences $\{f_\alpha; \alpha \in \Lambda\}$, where convergence is understood with respect to the metric induced by the scalar product in Eq.~\eqref{eqn::scalar}. To put it more intuitively, the space $\Fcal^0$ is the space spanned by all infinite tensor products $\bigotimes_{\alpha \in \Lambda} f_\alpha$ which have a finite norm, and $\Hcal_\Lambda$ is its completion. On this Hilbert space, we can define the space of trace class operators $\Bcal_1(\Hcal_\Lambda)$, and we shall denote the complex vector space spanned by CP maps $\Mcal: \Bcal_1(\Hcal_\Lambda) \rightarrow \Bcal_1(\Hcal_\Lambda)$ by $\Lcal_\Lambda$. 

Intuitively, the general stochastic process $\Tcal_\Lambda$, whose existence we will prove below, should be a functional of the form $\Tcal_\Lambda: \Lcal_\Lambda \rightarrow \mathbbm{C}$.
However, as we will see, the space $\Lcal_\Lambda$ is slightly `too big' for two distinct reasons, and $\Tcal_\Lambda$ will only be uniquely defined on a smaller, naturally arising space. With these preliminary definitions of the involved spaces out of the way, we can now prove the first part of the theorem: 

\textbf{(i)} \textbf{Existence of a unique comb $\Tcal_\Lambda^\sharp$ on the set $\Lambda$.}

Let $\Lambda$ be a (possibly uncountable) set, $\left\{\Lambda_k\right\}_{\Lambda_k\subseteq \Lambda}$ the set of all finite subsets of $\Lambda$, and let $\left\{\Tcal_{\Lambda_k}\right\}_{\Lambda_k\subseteq \Lambda}$ be the corresponding family of combs.  Consequently, we have $\Tcal_{\Lambda_k}: \Lcal_{\Lambda_k} \rightarrow \mathbbm{C}$. Now, let the family of combs satisfy the consistency condition of Eq.~\eqref{eqn::Consistency}
for all finite $\Lambda_k \subseteq \Lambda_\ell \subseteq \Lambda$. 

With this, we can `lift' the family of combs to a comb $\Tcal_{\Lambda}^\sharp$ that acts on the full space $\Lcal_\Lambda$ and contains all of the finite combs as `marginals'. To this end, we first define the inverse projection $\pi_{\Lambda_k}^{-1}: \Lcal_{\Lambda_k} \rightarrow \Lcal_{\Lambda}$ which trivially extends every map $\xi_{\Lambda_k} \in \Lcal_{\Lambda_k}$ that is only defined on a finite number of times to an operator 
\begin{gather}
\label{eqn::InvProj}
    \pi_{\Lambda_k}^{-1}[\xi_{\Lambda_k}] = \xi_{\Lambda_k}\otimes  \bigotimes_{t_\alpha \in \Lambda\setminus \Lambda_k} \Ical_{t_\alpha}
\end{gather}
that is defined on all times.\footnote{We will denote elements of $\Lcal_{\Lambda_k}$ by $\xi_{\Lambda_k}$ instead of $\Mcal_{\Lambda_k}$ to emphasize that they are not necessarily CP maps, but rather complex linear combinations of CP maps.} In other words, $\pi_{\Lambda_k}^{-1}$ maps any $\xi_{\Lambda_k}$ to a corresponding operator that lies in $\Lcal_{\Lambda}$ and only acts non-trivially on $\Lcal_{\Lambda_k}$. The operator $\pi_{\Lambda_k}^{-1}[\xi_{\Lambda_k}]$ exists and is unique for any finite $\Lambda_k \in \Lambda$ and all $\xi_{\Lambda_k} \in \Lcal_{\Lambda_k}$~\cite{neumann_infinite_1939}. 

In the same way, we can define a partial inverse projection $\pi^{-1}_{\Lambda_k \leftarrow \Lambda_\ell}: \Lcal_{\Lambda_k} \rightarrow \Lcal_{\Lambda_\ell}$ for any two finite sets $\Lambda_k \subseteq \Lambda_\ell \subseteq \Lambda$, \textit{i.e.}, 
\begin{gather}
\label{eqn::InvProjPart}
    \pi^{-1}_{\Lambda_k \leftarrow \Lambda_\ell}[\xi_{\Lambda_k}] = \xi_{\Lambda_k} \otimes \bigotimes_{t_\alpha \in \Lambda_\ell\setminus \Lambda_k} \Ical_{t_{\alpha}}\, .
\end{gather}
Employing these partial inverse projections, the consistency property of the family $\{\Tcal_{\Lambda_k}\}$ reads 
\begin{gather}
\label{eqn::Consistency_proj}
    \Tcal_{\Lambda_k}[\xi_{\Lambda_k}] = \Tcal_{\Lambda_\ell}\left[\pi^{-1}_{\Lambda_k \leftarrow \Lambda_\ell}[\xi_{\Lambda_k}]\right].
\end{gather}
All of the lifted operators $\pi^{-1}_{\Lambda_k}[\xi_{\Lambda_k}]$ are elements of $\Lcal_\Lambda$. Let $\Lcal_{\Lambda}^\sharp \subset \Lcal_\Lambda$ denote the set of all lifted operators, \textit{i.e.}, for all $\Lambda_k$ finite, we have
\begin{gather}
\mathcal{L}_{\Lambda}^\sharp = \{\xi \in \Lcal_\Lambda| \xi = \xi_{\Lambda_k} \otimes \bigotimes_{t_\alpha \in \Lambda \setminus \Lambda_k} \Ical_{t_\alpha}\}\, .
\end{gather}
On this space, we can define a comb $\Tcal_{\Lambda}^\sharp$ via 
\begin{gather}
\label{eqn::Tsharp}
    \Tcal_{\Lambda}^\sharp[\xi] := \Tcal_{\Lambda_k}[\xi_{\Lambda_k}]\,, \,\mathrm{ where } \ \xi = \pi_{\Lambda_k}^{-1}[\xi_{\Lambda_k}]\,.
\end{gather}  
It remains to show that $\Tcal_{\Lambda}^\sharp$ is well-defined in the sense that it maps every $\xi \in \Lcal_\Lambda^\sharp$ to a unique value. Specifically, if there are two different operators $\xi_{\Lambda_k}\in \Lcal_{\Lambda_k}$ and $\xi_{\Lambda_\ell}\in \Lcal_{\Lambda_\ell}$ that are lifted to the same $\xi \in \Lcal_\Lambda$, such that 
\begin{gather}
\label{eqn::uniqueness}
    \xi = \pi_{\Lambda_k}^{-1}[\xi_{\Lambda_k}]= \pi_{\Lambda_\ell}^{-1}[\xi_{\Lambda_\ell}]\, ,
\end{gather}
then $\Tcal_{\Lambda}^\sharp[\xi] = \Tcal_{\Lambda_k}[\xi_{\Lambda_k}]$ but also $\Tcal_{\Lambda}^\sharp[\xi] = \Tcal_{\Lambda_\ell}[\xi_{\Lambda_\ell}]$ and $\Tcal_{\Lambda}^\sharp$ might not be well-defined. However, uniqueness of $\Tcal_{\Lambda}^\sharp[\xi]$ is ensured by the consistency property; from Eq.~\eqref{eqn::uniqueness}, it is straightforward to see that 
\begin{align}
\notag
    \pi_{\Lambda_k \leftarrow \Lambda_k\cup \Lambda_\ell}^{-1}[\xi_{\Lambda_k}] &= \pi_{{\Lambda_\ell} \leftarrow \Lambda_k\cup \Lambda_\ell}^{-1}[\xi_{\Lambda_\ell}] \\
    &=: \xi_{\Lambda_k\cup\Lambda_\ell}\, .
\end{align}
Employing the consistency condition~\eqref{eqn::Consistency_proj} then yields 
\begin{align}
\notag
\Tcal_{\Lambda_k}&[\xi_{\Lambda_k}] = \Tcal_{\Lambda_k \cup \Lambda_\ell}[\pi_{\Lambda_k \leftarrow \Lambda_k \cup \Lambda_\ell}^{-1}[\xi_{\Lambda_k}]] \\ \notag &= \Tcal_{\Lambda_k \cup \Lambda_\ell}[\xi_{\Lambda_k \cup \Lambda_\ell}] \\ &=  \Tcal_{\Lambda_k \cup \Lambda_\ell}[\pi_{\Lambda_\ell \leftarrow \Lambda_k\cup \Lambda_\ell}^{-1}[\xi_{\Lambda_\ell}]] = \Tcal_{\Lambda_\ell}[\xi_{\Lambda_\ell}]\, .
\end{align}
Consequently, $\Tcal_{\Lambda}^\sharp[\xi]$ is independent of the representation of $\xi$. Additionally, by construction we have $\Tcal_\Lambda^{\sharp|\Lambda_k} = \Tcal_{\Lambda_k}$, \textit{i.e.}, $\Tcal_\Lambda^\sharp$ contains all finite combs of the family $\{\Tcal_{\Lambda_k}\}$ as `marginals'. However, so far  $\Tcal_\Lambda^\sharp$ is only defined on the set $\Lcal_\Lambda^\sharp \subset \Lcal_\Lambda$. In the second part of the proof, we show that $\Tcal_\Lambda^\sharp$ can be uniquely extended to a linear functional on a bigger space $\overline{\Lcal}_{\Lambda}^{\sharp} \supseteq \Lcal_{\Lambda}^{\sharp}$. 

\textbf{(ii) Extension of $\Tcal_{\Lambda}^\sharp$.}

In order to extend $\Tcal_{\Lambda}^\sharp$ to a linear functional $\overline{\Tcal}_{\Lambda}^\sharp$ that acts on elements of (a subest of) the closure $\overline{\Lcal}_\Lambda^\sharp$ of $\Lcal_\Lambda^\sharp$ , we will make use of the fact that any linear bounded mapping from a subset $\Kcal_{\Lambda}^\sharp$ of a normed vector space $X$ to a normed complete vector space $Y$ can be uniquely extended to a linear transformation from the completion $\overline{\Kcal}_{\Lambda}^\sharp$ of $\Kcal_{\Lambda}^\sharp$ to $Y$ (see, \textit{e.g.}, Thm. $2.7$-$11$ of ~\cite{kreyszig_introductory_1989}). 

So far, we have considered $\Tcal_{\Lambda}^\sharp$ as a mapping $\Tcal_{\Lambda}^\sharp: \Lcal_{\Lambda}^{\sharp} \rightarrow \mathbbm{C}$. However, $\Tcal_{\Lambda}^\sharp$ is not necessarily bounded on $\Lcal_{\Lambda}^{\sharp}$; as we discussed in Sec.~\ref{subsec::GET}, the action of a $k$-step comb $\Tcal_{\Lambda_k}$ is only meaningfully defined on the set $\Kcal_{\Lambda_k}$ of maps that agree with the causal ordering of $\Tcal_{\Lambda_k}$. In general, there will be many  maps in $\Lcal_{\Lambda_k}$ that have, for example, an opposite causal ordering than the one $\Tcal_{\Lambda_k}$ abides by. Thus, the action of $\Tcal_{\Lambda_k}$ on such a map would create causal loops and lead to `probabilities' that exceed $1$. The norm of $\Tcal_{\Lambda_k}$ on $\Lcal_{\Lambda_k}$ would thus depend on the number of possible causal loops, which, in turn, depends on the number of times in $\Lambda_k$. This, finally, implies that, in principle, for every positive number $r$, we could find a CPTP map $\Mcal^r\in \Lcal^\sharp_\Lambda$ with unit norm, such that $\Tcal_\Lambda^\sharp[\Mcal^r] > r$, rendering $\Tcal_\Lambda^\sharp$ unbounded on $\Lcal_{\Lambda}^\sharp$. While this unboundedness is not a problem for the first part of the proof, it keeps us from uniquely extending $\Tcal_{\Lambda}^\sharp$ to a linear functional on the closure of $\Lcal_\Lambda^\sharp$.

However, this is not a conceptual problem, as a CPTP map $\Mcal^r$ that yields a probability higher than $1$ cannot be implemented within the causal order the combs $\Tcal_{\Lambda_k}$ abide by. Consequently, without losing generality, we can restrict ourselves to the respective subsets of operations that the combs $\Tcal_{\Lambda_k}$ are meaningfully defined on. Specifically, we set 
\begin{align}
\notag
    \Kcal_{\Lambda}^\sharp := \{\xi \in \Lcal^\sharp_\Lambda&| \xi = \xi_{\Lambda_k} \otimes \bigotimes_{t_\alpha \in \Lambda \setminus \Lambda_k} \Ical_{t_\alpha}, \\
    &\phantom{|}\  \xi_{\Lambda_k}\in \Kcal_{\Lambda_k} \}\, , 
\end{align} 
\textit{i.e.}, $\Kcal_{\Lambda}^\sharp$ contains all trivial extensions of operators that the finite combs $\Tcal_{\Lambda_k}$ are meaningfully defined on. As $\Kcal_{\Lambda}^\sharp \subset \Lcal_{\Lambda}^\sharp$, $\Tcal_\Lambda^\sharp$ is uniquely defined on $\Kcal_{\Lambda}^\sharp$. Now, considering the mapping 
\begin{gather}
    \Tcal_{\Lambda}^\sharp: \Kcal_{\Lambda}^\sharp \rightarrow \mathbbm{C}\, ,
\end{gather}
we can show that there exists a unique extension $\overline{\Tcal}_{\Lambda}^\sharp: \overline{\Kcal}_{\Lambda}^\sharp \rightarrow \mathbbm{C}$ that has the desired properties. To this end, we have to show that $\Lcal_{\Lambda}^{\sharp}$ is a normed vector space, and that $\Tcal_{\Lambda}^\sharp$ is linear, and bounded on $\Kcal_{\Lambda}^\sharp$ (the space $\mathbbm{C}$ is well-known to be a complete vector space). We start with the former: 

Let $\beta, \gamma \in \mathbbm{C}$ and $\xi = \pi_{\Lambda_k}^{-1}[\xi_{\Lambda_k}], \zeta = \pi_{\Lambda_\ell}^{-1}[\zeta_{\Lambda_\ell}]$. It follows directly from the definition~\eqref{eqn::InvProj} of the inverse projection that
\begin{gather}
     \beta\, \xi = \pi_{\Lambda_k}^{-1}[\beta\, \xi_{\Lambda_k}] \in \Lcal_{\Lambda}^{\sharp}
\end{gather}
Now, we define  $\Gamma_{\Lambda_k\cup \Lambda_\ell} \in \Lcal_{\Lambda_k\cup\Lambda_\ell}$ as
\begin{align}
    \notag
    &\Gamma_{\Lambda_k\cup \Lambda_\ell} \\
    &= \pi^{-1}_{\Lambda_k \leftarrow \Lambda_k\cup\Lambda_\ell}[\beta \, \xi_{\Lambda_k}] + \pi^{-1}_{\Lambda_\ell \leftarrow \Lambda_k\cup\Lambda_\ell}[\gamma \, \zeta_{\Lambda_\ell}] .
\end{align}
One immediately sees that
\begin{gather}
\beta \, \xi + \gamma \, \zeta = \pi^{-1}_{\Lambda_k\cup \Lambda_\ell}[\Gamma_{\Lambda_k\cup \Lambda_\ell}] \in \Lcal_\Lambda^\sharp\, ,
\end{gather} 
which implies that $\Lcal_{\Lambda}^{\sharp}$ is a complex vector space. Additionally, $\mathcal{L}_{\Lambda}^\sharp$ becomes a \textit{normed} vector space by setting $\| \xi \| = \| \pi_{\Lambda_k}^{-1}[\xi_{\Lambda_k}] \| := \| \xi_{\Lambda_k} \|_{\mathrm{op}}$, where $\|\cdot\|_{\mathrm{op}}$ is the norm on $\Lcal_{\Lambda_k}$ induced by the trace norm on $\Bcal_1(\Hcal_{\Lambda_k})$, \textit{i.e.}, 
\begin{gather}
    \| \xi_{\Lambda_k} \|_{\mathrm{op}} = \sup_{|A|_{\mathrm{tr}} = 1} \{|\xi[A]|_\mathrm{tr}\ |\ A \in \Bcal_1(\Hcal_{\Lambda_k})\}\, .
\end{gather}

To prove linearity, and boundedness on $\Kcal_\Lambda^\sharp$ of $\Tcal_{\Lambda}^\sharp$, we make use of the linearity and boundedness of the finite combs $\Tcal_{\Lambda_k}$: For all $\Lambda_k$, we have $\Tcal_{\Lambda_k}[\Mcal_{\Lambda_k}] \leq 1$ for all $\Mcal_{\Lambda_k} \in \Kcal_{\Lambda_k}$. As this bound is uniform, \textit{i.e.}, independent of the set of times $\Lambda_k$, we immediately see that $\Tcal_{\Lambda}^\sharp[\Mcal_\Lambda] \leq 1$ for all $\Mcal_\Lambda \in \Kcal_\Lambda^\sharp$, rendering $\Tcal_{\Lambda}^\sharp$ bounded on $\Kcal_\Lambda^\sharp$.

The linearity of $\Tcal_{\Lambda}^\sharp$ follows in a similar vein; due to the linearity of $\Tcal_{\Lambda_k}$ and the linearity of the inverse projection operators, for all $ \beta, \gamma \in \mathbbm{C}$ and all $\eta, \xi \in \Lcal_{\Lambda}^\sharp$ we have 
\begin{gather}
    \Tcal_{\Lambda}^\sharp[\beta \xi + \gamma \eta] = \beta \Tcal_{\Lambda}^\sharp[\xi]  + \gamma \Tcal_{\Lambda}^\sharp[\eta]\,. 
\end{gather}
Consequently, there exists a unique comb $\overline{\Tcal}^\sharp_{\Lambda}$ defined on the completion $\overline{\Kcal}_{\Lambda}^{\sharp}$ of $\Kcal_{\Lambda}^{\sharp}$ that has, by construction, the family $\left\{\Tcal_{\Lambda_k} \right\}_{ \Lambda_k\subseteq \Lambda}$ as `marginals'. As $\Tcal_{\Lambda}^\sharp$ is positive and bounded by $1$ on $\Kcal_\Lambda^\sharp$, by continuity so is $\overline{\Tcal}_{\Lambda}^\sharp$ on $\overline{\Kcal}^\sharp_\Lambda$. This concludes the proof. 
\end{proof}

The space $\overline{\Kcal}_\Lambda^\sharp$ is a proper subset of $\overline{\Lcal}_\Lambda^\sharp$ (this latter space is sometimes called \textit{quasilocal algebra} in the literature~\cite{haag_algebraic_1964,kretschmann_quantum_2005}). There are two important points to note about these spaces. On the one hand, it is clear that in most relevant cases,  $\overline{\Kcal}_\Lambda^\sharp$ cannot coincide with $\overline{\Lcal}_\Lambda^\sharp$. As mentioned above, $\Lcal_\Lambda^\sharp$ generically contains CPTP maps that do not agree with the causal order of the finite combs $\Tcal_{\Lambda_k}$, and extending $\Tcal_{\Lambda}^\sharp$ to a linear functional on $\overline{\Lcal}_\Lambda^\sharp$ would neither be possible (as outlined above) nor meaningful. In this sense, $\overline{\Kcal}_\Lambda^\sharp$ is the `biggest' space that we can extend the finite functionals to, and it can be understood as the set of \textit{all} operations on times in $\Lambda$ that abide with the causal order of the finite combs $\Lambda_k$. As such, it is not just the `biggest' space we can extend the action of $\Tcal_{\Lambda}^\sharp$ to, but also the most meaningful one.

On the other hand, it is important to note that the space $\overline{\Lcal}_{\Lambda}^{\sharp}$ does \textit{not} coincide with $\Lcal_{\Lambda}$ (they coincide iff  $\Lambda$ is finite~\cite{neumann_infinite_1939}). Consequently, there might be different combs $\Tcal_\Lambda$ defined on $\Lcal_{\Lambda}$ with coinciding action on all elements of $\overline{\Kcal}_\Lambda^\sharp$. This, however, is not problematic; first, $\overline{\Lcal}_\Lambda^\sharp$ ``is in a way more important than'' $\Lcal_{\Lambda}$ because its elements arise from the ones of $\Lcal(\Hcal_{\alpha})$ ``by extension and algebraical and topological processes''~\cite{neumann_infinite_1939}. Second, just as for the KET~\cite{breuer_theory_2007}, the different possible combs on $\Lcal_\Lambda$ all lead to the same measurement statistics on any experimentally accessible set of times, so this non-uniqueness is not accessible/detectable in practice.

We emphasize that, even though we have phrased the above in the language of quantum mechanics, there is nothing particularly quantum mechanical about the GET. The proof of the theorem uses the compatibility, linearity and boundedness of the combs $\Tcal_{\Lambda_k}$, as well as the assumption that the spaces they act on span a vector space. Consequently, it holds for \textit{any} probabilistic theory (with interventions) satisfying these minimal assumptions.

Furthermore, the input and output spaces of the CP maps the comb acts on do not have to be of the same dimension. In this case, the identity map used for the consistency condition has to be slightly generalized: A CPTP map $\Mcal_{\alpha}: \Bcal_1(\Hcal_\alpha^{\inp}) \rightarrow \Bcal_1(\Hcal_{\alpha}^{\out})$ is implemented via a corresponding unitary $U_{\alpha}$, a \textit{fixed} ancillary state $\eta_\alpha \in \Bcal_1(\Hcal_{\mathrm{A}_\alpha})$, and a partial trace $\tr_{\mathrm{B}_\alpha}$ that is such that the resulting state $\Mcal_{\alpha}[\rho] = \tr_{\mathrm{B}_\alpha}\left[U_{\alpha}\left(\rho \otimes \eta_\alpha\right)U_{\alpha}^{\dagger}\right]$ lies in $\Bcal_1(\Hcal_{\alpha}^{\out})$. With this, we can define a \textit{generalized identity map} 
\begin{gather}
    \Ical_{t_\alpha}^{(\inp\rightarrow \out)}[\rho] = \tr_{\mathrm{B}_\alpha}\left(\rho \otimes \eta_\alpha\right),
\end{gather} and the GET still holds. The only difference being that the inverse projections used in its derivation, and given in Eqs.~\eqref{eqn::InvProj} and~\eqref{eqn::InvProjPart},  have to be changed accordingly to account for the altered identity map. Consequently, our theorem accounts for the case where particles are created/annihilated in the process, as well as the case where different degrees of freedom are manipulated at each time $t_\alpha$, or where the number of measurement outcomes and active interventions differ. 

Even more generally, the particular form of the `do nothing' operation, \textit{i.e.}, the action on the system of interest in the absence of active experimental intervention is of no importance for the derivation of the GET. In case that it does not coincide with the identity map $\Ical$ (or the more general identity map discussed above) but is represented by some map $\Mcal$ (for example, one could imagine a theory where nature constantly measures the system of interest) the logic of the proof of the GET would still follow through. Again, the only change in its derivation would be an adjustment of the inverse projection maps of Eqs.~\eqref{eqn::InvProj} and~\eqref{eqn::InvProjPart}, with the rest of the proof staying the same.

In the derivation of the GET, we make the implicit assumption that the employed CP maps only depend on the measurement outcome they correspond to, but not on the particular instrument that was used to carry out the respective measurement. This property has been dubbed `instrument non-contextuality'~\cite{caves_gleason-type_2004,shrapnel_updating_2017} or `operational instrument equivalence'~\cite{shrapnel_causation_2018}. In principle, our derivation could be straightforwardly adapted to any theory, where this assumption is no longer satisfied, but in which probabilities are still a linear function of the maps \textit{and} their respective contexts (\textit{i.e.}, instruments). Instead of the identity map, one would then use a pair $(\Ical,\Jcal_\Ical)$ of identity map and identity context for marginalization, and the GET would still hold.

It is important to clearly distinguish between the classical Kolmogorov extension theorem and the GET. The KET hinges on the fact that, in classical physics, a measurement does not change the average state of a system. This fails to hold in quantum mechanics, or any theory with interventions.  More concretely, in the language of quantum maps, the sum over the outcomes $i_\alpha$ of a measurement in a basis $\{\ket{i}\}$ at time $t_\alpha$ corresponds to the CPTP map $\Mcal_\alpha = \sum_{i_\alpha}\Mcal_{i_\alpha}$, where $\Mcal_{i_\alpha}[\rho] = \bra{i_\alpha}\rho\ket{i_\alpha} \ketbra{i_\alpha}{i_\alpha}$. In a classical stochastic process, the state $\rho$ is diagonal in the basis $\{\ket{i}\}$, and we have $\Mcal_\alpha[\rho] = \rho$; the average over measurement outcomes has the same effect as the classical `do nothing' operation. As soon as $\rho$ is not diagonal in the measurement basis, we have $\Mcal_{\alpha}[\rho]\neq \rho$; on average, a measurement in quantum mechanics changes the state of the system and the future measurement statistics will depend on the measurement that was performed. Consequently, joint probability distributions in classical physics (without interventions) exhibit a consistency condition, while quantum processes (and theories with interventions) generally do not. 

As in the classical case, the proof of the GET does not assume an \textit{a priori} temporal ordering. The sets $\Lambda_k$ could be sets of times, but also labels of different laboratories. We have the following remark:

\begin{remark*}
The proof of the GET does \textit{not} assume any ordering of the sets $\Lambda_k$, and only uses the generalized consistency property of Eq.~\eqref{eqn::Consistency} as its main ingredient. 
\end{remark*}

As alluded to above, this implies that the GET also applies to causally indefinite processes~\cite{PhysRevA.88.022318, OreshkovETAL2012, araujo_witnessing_2015}, as the descriptors for different sets of laboratories would still satisfy a compatibility condition. However, these processes do not admit a Stinespring dilation that is compatible with a fixed causal order~\cite{PhysRevA.88.022318, OreshkovETAL2012} and the interpretation of an underlying `process' becomes much less clear in the absence of a definitive causal ordering. We will briefly remark on this further in our conclusions, but leave a full exploration of this interpretation as an open question for future work. Next, we will see that the distinction between stochastic processes and causal modelling does not exist in the general case.

\subsection{Quantum stochastic processes and quantum causal modelling}
\label{subsec::QSP_QCM}
Using an instrument at some intermediate time $t_\alpha$ alters the state of a quantum system (even when averaging over all outcomes) and influences the statistics of later measurements in a non-negligible way. Nevertheless, the \textit{full} descriptor of an $\ell$-step process, \textit{i.e.}, $\Tcal_{\Lambda_\ell}$, contains all descriptors $\Tcal_{\Lambda_k}$ for fewer times $\Lambda_k \subseteq \Lambda_\ell$, and a family of compatible combs implies the existence of an underlying stochastic process $\Tcal_\Lambda$.

Like in the classical case, the GET provides the mathematical underpinnings for the theory of stochastic processes in quantum mechanics, or any other theory with interventions, and fixes the minimal necessary requirements for the existence of an underlying process. As we have seen, in quantum mechanics, it is unavoidable to employ a description that takes interventions into account, when attempting to obtain a consistent description of a quantum process; if one wants to properly define quantum stochastic processes, one is forced to use the framework of causal modelling where active interventions are used to infer the causal relations between different events. This motivates the following observation:

\begin{observation}
The theory of quantum causal modelling and the theory of quantum stochastic processes are equivalent.
\end{observation}

In contrast to Observation~\ref{lem::ccm_csp}, the set of quantum causal models does not just contain the set of quantum stochastic processes but coincides with it; in classical physics, we obtain a consistent description of stochastic processes without taking interventions into account, and we can \textit{choose} to intervene whenever we want to probe the causal structure of a process. In quantum mechanics, a consistent description of stochastic processes can \textit{only} be recovered if interventions are included in the description from the start. Interventions are not a choice but a necessity in quantum mechanics, which leads to the equivalence of quantum causal modelling and quantum stochastic processes. 

This implies that the breakdown of the KET in quantum mechanics is \textit{fundamental}, while it can in principle be removed by changing perspective in a classical process with interventions. In the latter case, a \textit{super-observer}, that observes both the experimenter manipulating the system of interest as well as the stochastic process itself, would obtain families of joint probability distributions that display a compatibility property. Put differently, for classical processes, by incorporating the experimenter and their choice of instrument into the stochastic process, the KET always applies on a higher level. In quantum mechanics, this is generally not true. No matter the level at which a super-observer observes a process, the respective joint probability distributions do not satisfy a compatibility property, and the KET fails to hold. This fundamental breakdown of the KET in quantum mechanics is mirrored by no-go theorems that show that non-contextual theories cannot reproduce the predictions of quantum mechanics; for many of these theorems, the notion of ontic latent variables~\cite{spekkens_contextuality_2005,cavalcanti_classical_2018} or ontic processes ~\cite{shrapnel_causation_2018} are introduced, and the basic assumption is made that the distributions over observable outcomes can be obtained by marginalization of a larger joint distribution over the values of the ontic variable. Subsequently, it is shown that, together with other assumptions, this prerequisite fails to reproduce predictions made by quantum mechanics. The GET dictates how to correctly compute marginals in quantum mechanics, such that all resulting probability distributions `fit together' and are the marginals of one common comb $\Tcal_\Lambda$. It is therefore conceivable that a derivation starting from the assumption of compatibility in the sense of the GET would lead to theories that can indeed reproduce quantum mechanics.

We reiterate that classical stochastic processes are a very special subset of general stochastic processes, namely the ones where the system of interest is never rotated out of its fixed (pointer) basis, and the experimenter can only perform projective measurements in this basis.\footnote{The set of quantum  processes that can be described by only classical means is in fact slightly bigger~\cite{strasberg_classical_2019,milz_when_2019}. We will comment on this subtlety below.} We now show that the KET can be derived in a straight forward way as a corollary of the GET.

\subsection{GET \texorpdfstring{$\Rightarrow$}{} KET} 
\label{subsec::GET_KET}
Our generalised extension theorem applies to a strictly larger class of theories than the standard KET and includes the latter as a corollary. Specifically, in the language introduced above, a classical process is one where the experimenter can only perform measurements in a fixed basis, and the resulting joint probability distributions satisfy Kolmogorov consistency conditions. With this -- under the aforementioned assumption that all considered value spaces are $\mathbbm{R}$, $\mathbbm{N}$, or, more generally, Borel spaces -- we have the following proposition:
\begin{proposition}\label{lem::getket}
The GET implies the KET.
\end{proposition}

\begin{proof}
In order to prove this statement, we will show that any family of classical joint probability distributions that satisfies the consistency property of the KET can be mapped onto a family of quantum combs that satisfies the consistency condition of the GET -- albeit with a slightly different identity map. The GET then guarantees that there exists an underlying classical comb $\Tcal^{\text{cl}}_{\Lambda}$, and thus also an underlying classical process $\mathbbm{P}_\Lambda$. 

Let $\{\Pprob_{\Lambda_k}\}_{\Lambda_k\subset \Lambda}$ be a family of joint probability distributions on all finite subsets of $\Lambda$ that satisfies the consistency conditions of the KET, \textit{i.e.}, $\Pprob_{\Lambda_k} = \Pprob_{\Lambda_\ell}^{|\Lambda_k}$ for all $\Lambda_k \subset \Lambda_\ell \subset \Lambda$. We denote the set of perfectly distinguishable possible outcomes at time $t_\alpha$ by $\{i_\alpha\}$. With this, we can define a Hilbert space $\Hcal_\alpha$ spanned by an orthogonal set of states $\{\ket{i_\alpha}\}$, and projective CP operators $\Pcal_{i_\alpha}$ that correspond to a measurement with outcome $i_\alpha$. The action of these operators on a state $\rho \in \Bcal_1(\Hcal_\alpha)$ is given by $\Pcal_{i_\alpha}[\rho] = \braket{i_\alpha| \rho |i_\alpha} \ketbra{i_\alpha}{i_\alpha}$. The complex vector space spanned by these projective operators will be denoted by $\Omega_\alpha$, and, correspondingly, we set $\Omega_{\Lambda_k} = \bigotimes_{\alpha \in \Lambda_k} \Omega_\alpha$. On said vector space, we can define a classical comb $\Tcal_{\Lambda_k}^{\mathrm{cl.}}$, with its action on every $\Pcal_{\mathbf{i}_{\Lambda_k}} = \bigotimes_{\alpha \in \Lambda_k} \Pcal_{i_\alpha}$ given by 
\begin{gather}
    \Tcal_{\Lambda_k}^{\mathrm{cl.}}[\Pcal_{\mathbf{i}_{\Lambda_k}}] = \Pprob_{\Lambda_k}({\mathbf{i}_{\Lambda_k}})\, .
\end{gather}
To stay closer in spirit to the proof of the GET, we could extend $\Tcal_{\Lambda_k}^{\mathrm{cl.}}$ to a CP linear functional on the whole space $\Lcal_{\Lambda_k} \supset \Omega_{\Lambda_k}$, but as this step is not necessary for the proof of the KET, we will not carry it out here.
As the family of probability distributions $\{\Pprob_{\Lambda_k}\}$ satisfies a consistency condition, we have for $\Lambda_k \subset \Lambda_\ell$
\begin{gather}
\sum_{\Lambda_\ell \setminus \Lambda_k} \Tcal_{\Lambda_\ell}^{\mathrm{cl.}}[\Pcal_{\mathbf{i}_{\Lambda_\ell}}] = \Pprob_{\Lambda_k}({\mathbf{i}_{\Lambda_k}}) = \Tcal_{\Lambda_k}^{\mathrm{cl.}}[\Pcal_{\mathbf{i}_{\Lambda_k}}]\, , 
\end{gather}
where $\mathbf{i}_{\Lambda_k}$ is the restriction of $\mathbf{i}_{\Lambda_\ell}$ to $\Lambda_k$. 
Setting $\Delta_{t_\alpha} := \sum_{i_\alpha} \Pcal_{i_\alpha}$, we see that the family of combs $\{\Tcal_{\Lambda_k}\}_{\Lambda_k \subset \Lambda}$ satisfies a consistency condition with respect to the operators $\Delta_{t_\alpha}$ (in contrast to the GET, where the corresponding operator  was $\Ical_{{t_\alpha}}$). As discussed above, the proof of the GET can be straightforwardly generalised to any choice of the `do-nothing' operation. Analogous to the proof of the GET, setting $\Omega_{\Lambda}^\sharp = \{\xi| \xi = \xi_{\Lambda_k} \otimes \bigotimes_{t_\alpha \in \Lambda \setminus \Lambda_k} \Delta_{t_\alpha}, \xi_{\Lambda_k} \in \Omega_{\Lambda_k}\}$ and
\begin{gather}
   \omega_\Lambda^\sharp = \{\xi \in \Omega_{\Lambda}^\sharp| \xi = \bigotimes_{\alpha \in \Lambda_k}  \Pcal_{i_\alpha} \otimes \bigotimes_{\Lambda\setminus \Lambda_k} \Delta_{t_\alpha}\} \, ,
\end{gather}
we see that there exists a unique comb $\overline{\Tcal}_{\Lambda}^{\mathrm{cl.}\sharp}$, defined on the closure $\overline{\omega}_{\Lambda}^\sharp$ of $\omega_{\Lambda}^\sharp$, that has the family $\{\Tcal_{\Lambda_k}^{\mathrm{cl.}}\}$ as marginals (with respect to the operators $\Delta_{t_\alpha}$).

It remains to show how to obtain a probability distribution $\Pprob_{\Lambda}$ from $\overline{\Tcal}_{\Lambda}^{\mathrm{cl.}\sharp}$ that contains all finite distributions as marginals. To this end, we note that the classical comb $\overline{\Tcal}_{\Lambda}^{\mathrm{cl.}\sharp}$ is well-defined and yields probabilities on all $\xi \in \overline{\omega}^\sharp \subset \overline{\Omega}^\sharp_\Lambda$. Every $\xi \in \overline {\omega}_\Lambda^\sharp \setminus \omega_\Lambda^\sharp$ that lies in the extension of $\omega_\Lambda^\sharp$ has a unique restriction $\xi_{|\Lambda_k} = \Pcal_{\mathbf{i}_{\Lambda_k}}$, to any \textit{finite} set $\Lambda_k$, and as such uniquely fixes a set of outcomes $\mathbf{i}_{\Lambda_k}$. Taking the union of all these sets of outcomes, we see that every $\xi \in \overline{\omega}^\sharp$ defines a set $\mathbf{i}_\Lambda$ (we will thus add an additional subscript and denote them by $\xi_{\mathbf{i}_\Lambda}$) of corresponding outcomes at all times in $\Lambda$. Setting $\Pprob_{\Lambda}(\mathbf{i}_\Lambda) = \overline{\Tcal}_{\Lambda}^{\mathrm{cl.}\sharp}[\xi_{\mathbf{i}_\Lambda}]$, we obtain a probability distribution $\Pprob_{\Lambda}$ that -- by construction -- yields the correct probability distributions $\Pprob_{\Lambda_k}$ when restricted to finite sets $\Lambda_k \subset \Lambda$.
\end{proof}

While the original version of the KET does not hold for quantum processes, it is important to note that the breakdown of the compatibility property of joint probability distributions is not a signature of quantum mechanics \textit{per se}; as we have already seen, any framework that allows for interventions will exhibit this feature. The GET provides a proper theoretical underpinning for the corresponding experimental situations. On the other hand, the breakdown of the compatibility property can happen in quantum mechanics even if only projective measurements in a fixed basis $\left\{\ket{i_\alpha}\right\}$ are performed~\cite{BreuerEA2016,smirne_coherence_2017}. 

As already mentioned, the absence of compatibility is tantamount to the absence of either realism \textit{per se}, or non-invasiveness (or both). Consequently, it can be used as a definition of non-classicality, as proposed in Ref.~\cite{smirne_coherence_2017}. There, the authors employ the breakdown of the consistency condition on the level of probability distributions, when measuring in a fixed basis, as a means to define the non-classicality of Markovian processes. Using the framework of quantum combs for the description of quantum stochastic processes the ideas of~\cite{smirne_coherence_2017} can be extended to general processes with memory, \textit{i.e.}, non-Markovian processes~\cite{strasberg_classical_2019, milz_when_2019}.

Following Ref.~\cite{smirne_coherence_2017}, we consider an $\ell$-step process to be classical if its joint probability distributions with respect to measurements in a fixed basis $\left\{\ket{i_\alpha}\right\}$ satisfy a consistency condition. Put differently, an $\ell$-step process $\Tcal_{\Lambda_k}$ is classical (with respect to the basis $\left\{\ket{i}\right\}$) iff for all $\Lambda_k \subseteq \Lambda_\ell$ \text{and} all possible sequences of outcomes $i_k,\dots, i_1$
\begin{gather}
\label{eqn::ClassProc}
    \Tcal_{\Lambda_k}[\Pcal_{i_k},\dots,\Pcal_{i_1}] = \sum_{\Lambda_\ell\setminus \Lambda_k} \Tcal_{\Lambda_\ell}[\mathcal{P}_{i_\ell},\dots,\mathcal{P}_{i_1}]\, , 
\end{gather}
where  $\Pcal_{i_\alpha}$ corresponds to obtaining outcome $i_\alpha$ from a projective measurement in a fixed basis at time $t_\alpha$, \textit{i.e.}, $\Pcal_{i_\alpha}[\rho] = \bra{i_\alpha}\rho\ket{i_\alpha}\ket{i_\alpha}\!\bra{i_\alpha}$.

The general structure of classical combs that satisfy Eq.~\eqref{eqn::ClassProc} can then be analyzed using the Choi isomorphism between quantum processes and positive matrices~\cite{Choi1975, jamiolkowski_linear_1972}. As combs can describe general processes with memory, Eq.~\eqref{eqn::ClassProc} represents a consistent
definition of classical processes with memory and allows a direct extension of the results obtained in Ref.~\cite{smirne_coherence_2017} to the non-Markovian case~\cite{strasberg_classical_2019,milz_when_2019}.

\section{Relation to previous works}
\label{sec::Relation}
As already mentioned, the proof of the GET does not rely on any particularities that are exclusive to quantum mechanics or our formulation thereof. The GET constitutes a sound basis for the description of \emph{any} conceivable (classical, quantum or beyond) theory of stochastic processes with interventions -- independent of the employed framework. 

While we referred throughout to the framework of \emph{quantum combs}~\cite{chiribella_transforming_2008, chiribella_quantum_2008, chiribella_theoretical_2009}, originally derived as the most general representation of quantum circuit architectures, our results apply equally well to any other framework for describing quantum processes as linear functionals. Examples of the mathematical objects and frameworks (often the same thing under a different name) given a firm theoretical foundation by the GET include: \emph{process tensors}~\cite{modi_operational_2012, pollock_non-markovian_2018, milz_introduction_2017} and \emph{causal automata/non-anticipatory channels}~\cite{kretschmann_quantum_2005, caruso_quantum_2014}, which describe the most general open quantum processes with memory; \emph{causal boxes}~\cite{portmann_causal_2015} that enter into quantum networks with modular elements; \emph{operator tensors}~\cite{hardy_operational_2016, hardy_operator_2012} and \emph{superdensity matrices}~\cite{cotler_superdensity_2017}, employed to investigate quantum information in general relativistic space-time; and, finally, \emph{process matrices}, used for quantum causal modelling~\cite{OreshkovETAL2012, 1367-2630-18-6-063032, oreshkov_causal_2016, allen_quantum_2017}. In classical physics, as well as the standard causal modelling framework discussed in Sec.~\ref{sec::ClassCaus}, our result applies to the  $\epsilon$-transducers used within the framework of computational mechanics~\cite{barnett_computational_2015,thompson_using_2017} to describe processes with active interventions. 

Our theorem proves the existence of a container space for all of the aforementioned frameworks and allows for their complete and consistent representation in the continuous time limit, thus providing an overarching theorem for probabilistic theories with interventions. This is of particular importance for the field of open quantum mechanics where the lack of an extension theorem has been a roadblock to obtaining a framework that coincides with classical descriptions in the correct limit~\cite{BreuerEA2016}. Here, switching perspective allows one to describe both classical as well as quantum open systems in a unified framework. This fact has recently been used to obtain an unambiguous definition of non-Markovianity in quantum mechanics that coincides with the classical one in the correct limit~\cite{pollock_operational_2018}.

The GET goes beyond previous attempts to generalize the KET for quantum mechanics. An extension theorem for positive operator valued measures was derived in Ref.~\cite{Tumulka2008} and was used in Ref.~\cite{haapasalo_saturation_2016} to show the existence of an `infinite composition' of an instrument. This extension theorem is, however, limited to particular cases of positive operator valued measures, and not general enough to provide an underpinning for the description of stochastic processes with interventions. 

More generally, a version of the KET for quantum processes was derived in Ref.~\cite{accardi_quantum_1982}. In this work, the authors showed that any quantum stochastic ``process can be reconstructed up to equivalence from a projective family of correlation kernels''. By decomposing the control operations $\Mcal_{i_\alpha}$ into their component Kraus operators, it can explicitly be shown that these correlation kernels correspond to combs, and consequently, for quantum processes, the GET is equivalent to Thm.~$1.3$ in Ref.~\cite{accardi_quantum_1982}. However, the mathematical structure of the latter does not tie in easily with recently developed frameworks for the description of quantum (or classical) causal modelling, nor does it lend itself in a straightforward way to the discussion of their key properties. Additionally, our proof -- in contrast to the one presented in~\cite{accardi_quantum_1982} -- highlights the role that causal order plays for the domain of the resulting stochastic process $\Tcal_\Lambda$. Specifically, while \textit{independent} CP maps at different times $t_i$ are considered in~\cite{accardi_quantum_1982}, our construction makes explicit the set of \textit{correlated} operations on $\Lambda$ that $\Tcal_{\Lambda}$ can be meaningfully applied to.

The structural features of combs render the investigation of fundamental features of a process, like their non-Markovianity~\cite{modi_operational_2012,pollock_operational_2018}, their causal structure~\cite{OreshkovETAL2012,1367-2630-18-6-063032,portmann_causal_2015}, and their classicality tractable. Furthermore, our formulation has the advantage that combs are defined in a clear-cut \textit{operational} way, and allow for a generalized Stinespring dilation~\cite{chiribella_theoretical_2009,pollock_non-markovian_2018}, which makes their interpretation in terms of open quantum system dynamics straightforward. Finally, even though the GET is stated for combs that map sequences of CP maps to probabilities, its proof also applies -- with slight modifications -- to general quantum combs (\textit{i.e.}, maps that map combs onto combs~\cite{chiribella_quantum_2008,chiribella_theoretical_2009}).

\section{Conclusions}
\label{sec::Concl}
While the KET is the fundamental building block for the theory of classical stochastic processes, it does not hold in quantum mechanics, or any other theory that allows for active interventions. This breakdown goes hand in hand with the violation of Leggett-Garg inequalities: the violation of such an inequality \textit{always} implies that compatibility conditions are not satisfied, and hence the KET does not hold.

In this work, we have proven a generalized extension theorem that applies to any process with interventions, including quantum ones. We have therefore shown that the roadblocks encountered when describing quantum processes in terms of joint probability distributions can be remedied by changing perspective; while the evolution of a density matrix over time does not contain enough statistical information for consistency properties to hold~\cite{BreuerEA2016}, considering a quantum stochastic process as a linear functional acting on sequences of CP maps allows one to formulate a fully fledged theory. Taking interventions into account is the only way to obtain a consistent definition and rigorous mathematical foundation for quantum stochastic processes. Put differently, without taking interventions into account, there is no way to consistently define quantum stochastic processes. In this sense, two seemingly different frameworks -- the framework of causal modelling, and the theory of quantum stochastic processes -- are actually two sides of the same coin.

In the limit of continuous time, the sequence of CP maps becomes a continuous driving/control of the system of interest. Thus, the GET provides the theoretical foundation for these experimental scenarios, important for development of quantum technologies. Likewise, just as in the case of classical stochastic processes, the GET provides a toolbox for the modelling of quantum stochastic processes; any mechanism that leads to consistent families of combs automatically defines an underlying process.

It is important to emphasize the generality of our main result. Due to the linearity of mixing, \textit{any} meaningful description of a stochastic process -- quantum or not -- must be expressible in terms of a linear function on the space of locally accessible operations~\cite{milz_introduction_2017}. The proof of the GET is versatile enough to account for any framework that aims to describe temporally ordered processes, and hence provides a sound mathematical underpinning for all of them. 

The GET contains the original KET as the special case where the family of processes is diagonal in the reference basis, and the only allowed CP maps are projective measurements in the same basis. On the one hand, this implies that our extension of classical processes to the quantum realm is the correct one. On the other hand, this clear-cut definition of classical combs lends itself ideally to the investigation of the interplay of coherence and classicality, as proposed in Ref.~\cite{smirne_coherence_2017}, in the experimental observation of real-world processes with memory.

Finally, our discussion made transparent where causality and causal order enter into the proof of the GET, and what sets of operations the resulting stochastic process can meaningfully be applied to. While we have mostly discussed temporally ordered processes, in principle, even causally disordered processes could be described by families of functionals that satisfy a consistency requirement ($\Lambda$ would then be thought of as a set of labels for different laboratories). However, there is no deterministic Stinespring dilation for causally disordered processes~\cite{PhysRevA.88.022318}. There are, on the other hand, dilations that include post-selection~\cite{chiribella_transforming_2008, milz_Entanglement_2017}, and we conjecture that an underlying causally disordered stochastic process would be equivalent to post-selection on a class of trajectories resulting from continuous weak measurement. 

\begin{acknowledgments}
\noindent We thank Andrea Collevecchio, Tim Garoni and Top Notoh for valuable discussions, and Roderich Tumulka for clarifying remarks on the KET for POVMs. We are especially grateful to the referees of this paper for providing thorough and constructive advice on how to improve the proof of the GET. KM and FAP thank an anonymous referee (of another paper) for pointing them to the lack of a KET in quantum mechanics. SM acknowledges funding from the the Monash International Postgraduate Research Scholarship (MIPRS), the J L William Scholarship,  the Austrian Science Fund (FWF): ZK3 (Zukunftkolleg), the  European  Unions  Horizon  2020  research and innovation programme under the Marie Sk{\l}odowska Curie grant agreement No 801110, and the Austrian Federal Ministry of Education, Science and Research (BMBWF). FS is grateful to thank Sri-Trang Thong Scholarship, Faculty of Science and ``Postgraduate Student Exchange Program 2016'', International Relations Division, Mahidol University, for financial support during his visit to Monash University, Australia. KM is supported through ARC FT160100073.
\end{acknowledgments}

\FloatBarrier


\bibliographystyle{apsrev4-1}
\bibliography{references}

\end{document}